\documentclass[12pt,a4paper]{article}

\usepackage{fixltx2e}
\usepackage{amsfonts} 

\usepackage[font=small]{caption}
\usepackage[text={450pt,650pt},headheight={14.5pt},centering]{geometry}

\usepackage{lmodern}
\usepackage{graphicx}
\usepackage{subfig}
\usepackage{booktabs}
\usepackage{multirow}
\usepackage{dsfont}
\usepackage{enumitem}
\usepackage{cite}
\usepackage{collref}

\usepackage{tikz}
\usetikzlibrary{plotmarks,calc,decorations.markings,decorations.pathmorphing,decorations.pathreplacing,patterns,matrix}

\tikzstyle arrow=[thick,rounded corners=18pt,-latex]
\tikzstyle box=[draw,rounded corners,outer sep=4pt]
\tikzstyle B node=[outer sep=0pt]
\tikzstyle Q node=[inner sep=1pt,outer sep=0pt]

\tikzset{%
  scalar/.style={thick,double},
  fermion/.style={thick,double,dashed},
  gluon/.style={thick,double,decorate,decoration={snake,amplitude=.25mm,segment length=1.25mm,post length=0.2mm,pre length=0.2mm}},
  hyp scalar/.style={thick},
  hyp fermion/.style={thick,dashed}
}

\newcommand{\BBt}{{\zeta}}
\newcommand{\BBr}{{\rho}}
\newcommand{\BBx}{{\xi}}

\usepackage[intlimits]{amsmath}
\usepackage{amssymb}
\usepackage{mathtools}
\usepackage{slashed}
\usepackage{braket}

\usepackage{hyperref}

\usepackage{xspace}

\usepackage{accents}

\numberwithin{equation}{section}

\makeatletter
\let\old@startsection=\@startsection
\let\oldl@section=\l@section
\renewcommand{\@startsection}[6]{\old@startsection{#1}{#2}{#3}{#4}{#5}{#6\mathversion{bold}}}
\renewcommand{\l@section}[2]{\oldl@section{\mathversion{bold}#1}{#2}}
\makeatother

\def\XXint#1#2#3{{\setbox0=\hbox{$#1{#2#3}{\int}$}
    \vcenter{\hbox{$#2#3$}}\kern-.5\wd0}}

\newcommand{\AdS}{\text{AdS}}
\newcommand{\CFT}{\text{CFT}}
\newcommand{\Sphere}{\mathrm{S}}
\newcommand{\Torus}{\mathrm{T}}
\newcommand{\CP}{\mathrm{CP}}

\newcommand{\alg}[1]{\mathrm{#1}}
\newcommand{\grp}[1]{\mathrm{#1}}

\newcommand{\grpSL}{\grp{SL}}

\newcommand{\algSU}{\alg{su}}

\newcommand{\grpSO}{\grp{SO}}

\newcommand{\grpO}{\grp{O}}

\newcommand{\grpU}{\grp{U}}

\newcommand{\Integers}{\mathbb{Z}}
\newcommand{\Reals}{\mathbb{R}}

\newcommand{\superN}{\mathcal{N}}

\newcommand{\eg}{\textit{e.g.}\xspace}

\newcommand{\pri}[1]{\accentset{\prime}{#1}}

\newcommand{\SymN}{\operatorname{Sym}_N}

\newcommand{\Vol}{\operatorname{vol}}

\begin{document}
\thispagestyle{empty}

\vspace*{2em}

\begin{center}
  \textbf{\Large\mathversion{bold} Closed Strings and Moduli in $\AdS_3/\CFT_2$}

  \vspace{2em}

  \textrm{\large 
    Olof Ohlsson Sax${}^1$ 
    and Bogdan Stefa\'nski, jr.${}^2$ 
  } 

  \vspace{4em}

  \begingroup\itshape
  ${}^1$Nordita, Stockholm University and KTH Royal Institute of Technology,\\
  Roslagstullsbacken 23, SE-106 91 Stockholm, Sweden\\[0.2cm]
  ${}^2$Centre for Mathematical Science, City, University of London,\\ Northampton Square, EC1V 0HB, London, UK\\[0.2cm]
  \par\endgroup

  \vspace{1em}

  \texttt{olof.ohlsson.sax@nordita.org, \\ Bogdan.Stefanski.1@city.ac.uk}


\end{center}

\vspace{3em}

\begin{abstract}\noindent
String theory on $\AdS_3\times\Sphere^3\times\Torus^4$ has 20 moduli.
We investigate how the perturbative closed string spectrum changes as we move around this moduli space in both the RR and NSNS flux backgrounds. We find that, at weak string coupling, only four of the moduli affect the energies. In the RR background the only effect of these moduli is to change the radius of curvature of the background. On the other hand, in the NSNS background, the moduli introduce worldsheet interactions which enable the use of integrability methods to solve the spectral problem. Our results show that the worldsheet theory is integrable across the 20 dimensional moduli space.
\end{abstract}

\newpage

\tableofcontents

\section{Introduction}

The $\AdS_3/\CFT_2$ correspondence~\cite{Maldacena:1997re} is an intriguing example of holography. One simple reason for this is the intermediate amount of supersymmetry, which allows rich and interesting physical phenomena, while at the same time retaining sufficient control for precise computations. For example, partly because of the low amount of supersymmetry, the dual pair has a large number of moduli. Perhaps a more significant reason for interest in this dual pair is the keystone role that $\AdS_3/\CFT_2$ plays in diverse areas, such as black-hole entropy~\cite{Strominger:1996sh}, the ADHM construction~\cite{Atiyah:1978ri} of instantons and their moduli space~\cite{Douglas:1996uz,Dijkgraaf:1998gf}, or holography of two-dimensional maximally-supersymmetric SQCD. For a review of the $\AdS_3/\CFT_2$ correspondence see~\cite{David:2002wn}.

A compelling  feature of  $\AdS_3/\CFT_2$ is that the D-brane construction used
to conjecture this duality does not directly provide a $\CFT_2$ living on the brane worldvolume. Rather, the low-energy dynamics of D1/D5 open strings is given by two-dimensional super-Yang-Mills theory (SYM) coupled to adjoint and fundamental matter. In the IR this theory flows to a direct sum of Coulomb and Higgs branch $\CFT$s~\cite{Witten:1997yu}, with the latter conjectured to be the dual to strings on the near-horizon $\AdS_3$ geometry~\cite{Maldacena:1997re}. This behaviour is in sharp contrast to the more supersymmetric cases where the D3- and M2-brane worldvolume gauge theories are conformal.

The lack of a perturbative gauge theory description has made direct tests of the $\AdS_3/\CFT_2$ correspondence more challenging. Based on topological arguments~\cite{Vafa:1995zh}, it has been suggested that the $\CFT_2$ should be closely related to a deformation of a $\SymN$ orbifold~\cite{Dijkgraaf:1998gf}. The location of the $\SymN$ orbifold point in the near-horizon moduli space has been discussed in~\cite{Larsen:1999uk}, but a precise understanding of the relationship between this orbifold and the full string theory remains to be understood more fully, as discussed for example on pp.~9-10 of~\cite{Seiberg:1999xz}. Nevertheless, many quantities protected by supersymmetry have been matched between strings on $\AdS_3$ and the $\SymN$ orbifold. However, these tests cannot establish the exact nature of the relationship between string theory on $\AdS_3$ and the $\SymN$ orbifold, precisely because the matching involves only protected quantitites and so does not depend on how, for example, the moduli enter the dictionary. One outstanding problem is to identify precisely  the states dual to perturbative closed strings in the $\SymN$ orbifold.

To make progress on this question, we investigate perturbative closed string states on $\AdS_3\times \Sphere^3\times\Torus^4$ with zero winding and zero momentum on $\Torus^4$, which we  denote by $\mathcal{H}_{(0,0)}$. Working in the Green-Schwarz formulation, we  determine what effect varying the 20 moduli has on the spectrum of $\mathcal{H}_{(0,0)}$. As is well known~\cite{Dijkgraaf:1998gf}, there are many type IIB string theory backgrounds on the space-time $\AdS_3\times \Sphere^3\times \Torus^4$. These differ from one another by the charges that the space-time carries. In this paper we will consider two such distinct $\AdS_3$ backgrounds corresponding to the near-horizon limit of D1/D5 and F1/NS5 brane configurations, returning to
backgrounds with more general charges in the future.\footnote{%
  In this paper, we will consider a string theory modulus to be a scalar field that can acquire a vacuum expectation value without changing the background charges. 
  This should be contrasted with a U-duality transformation which will generically change the charges of the background. 
  For example, changing one of the radii of $\Torus^4$ is a modulus of the D1/D5-brane or the F1/NS5-brane near-horizon theories. On the other
  hand, performing an S-duality maps one between the D1/D5 and F1/NS5 backgrounds and so should not be thought of as a marginal deformation.
} %
We find that for both types of $\AdS_3$ backgrounds only 4 moduli affect the $\mathcal{H}_{(0,0)}$ spectrum. In the D1/D5 background, these 4 are the dilaton $\phi$ and the 3 moduli associated with the self-dual NSNS potential $B^+$. In the F1/NS5 brane background these are 
instead the 3 moduli associated to the self-dual part of the RR two-form potential $C_2^+$ on $\Torus^4$ and  the modulus associated with a linear combination of $C_0$ and $C_4$.

In the $\AdS_3$ theory supported by RR flux, backgrounds with different values of $B^+$ are related by T-duality. As a result, the energies of $\mathcal{H}_{(0,0)}$ states are unchanged apart from a modification of the radius of curvature of the theory
\begin{equation}
  \label{eq:rad-d1d5-bmod}
  R^2=\alpha' g_s  N_5\sqrt{1+\tfrac{1}{2}(B^+)^2},
\end{equation}
where $N_5$ is the number of D5-branes of the background. Consequently, energies of $\mathcal{H}_{(0,0)}$ states at any value of these moduli can be determined using the Bethe Equations (BEs) found in~\cite{Borsato:2016xns} by simply modifying $R$ in this way.

Turning on the four RR moduli in the $\AdS_3$ theory supported by NSNS flux has a more significant effect on the worldsheet action. As we show, turning on these moduli induces non-zero RR field-strength couplings and the world-sheet action takes the same form (up to T-dualities) as the mixed-flux theories 
considered in~\cite{Cagnazzo:2012se,Hoare:2013ida,Lloyd:2014bsa}. The energies of $\mathcal{H}_{(0,0)}$ states then follow from the BEs found in~\cite{Borsato:2016xns}, using the exact S matrices determined in~\cite{Hoare:2013ida,Lloyd:2014bsa}, upon a suitable identification of parameters. We show that in the $\AdS_3$ theory supported by NSNS flux the strength of the integrable interactions, conventionally denoted by $h$ is given by\footnote{In the integrable literature this is conventionally denoted by $h(\lambda)$ where $\lambda=R^4/\alpha'{}^2$ is the 't Hooft coupling constant. In this paper, we will write $h(R)$ rather than $h(\lambda)$, since $R$ is a more natural parameter from the string theory point of view.}
\begin{equation}
h=-\frac{g_s c_0k}{2\pi}+\dots,
\end{equation}
where $k$ is the WZW level, $c_0$ is the value of the RR moduli, and subleading corrections in $R$ are denoted by ellipses. Setting the RR moduli to zero turns off the integrable interactions, which reduces the model to a GS version of the WZW $\CFT$ analysed by Maldacena and Ooguri~\cite{Maldacena:2000hw}. What is more, by analogy with higher-dimensional holography we see that (at large $R$) $c_0 k$ plays the role of the 't Hooft coupling constant $\sqrt{\lambda}$, and one is led to considering suitable double-scaling limits of $c_0$ and $k$ to capture leading (planar) dynamics.

This paper is organised as follows. In section~\ref{sec:moduli} we review the moduli space of strings on $\AdS_3\times \Sphere^3\times\Torus^4$. In sections~\ref{sec:d1-d5-moduli} and~\ref{sec:f1-ns5-moduli} we determine the effect of moduli on $\mathcal{H}_{(0,0)}$ in the D1/D5 and F1/NS5 backgrounds, respectively. In section~\ref{sec:backreacted} we present the complete backreacted geometries with non-zero moduli. We conclude in section~\ref{sec:conclusions}. The more technical aspects of our results are relegated to the appendices.

\section{The Moduli}
\label{sec:moduli}

In this section we briefly review the moduli space of Type IIB strings on $\AdS_3\times \Sphere^3\times\Torus^4$, summarising some of the results of~\cite{Larsen:1999uk}. Type IIB compactified on $\Torus^4$, in directions $x^6,\dots,x^9$, has 25 scalars which parametrise the coset $\frac{\grpSO(5,5)}{\grpSO(5)\times \grpSO(5)}$.
By wrapping D5-branes (respectively, NS5-branes) on $\Torus^4$, and D1-branes (fundamental strings)  transverse to $\Torus^4$ but parallel to the other 5-brane directions we obtain the BPS D1/D5 (F1/NS5) background. In the near-horizon limit of both geometries, five of the above scalars get fixed through the attractor mechanism~\cite{Ferrara:1995ih,Ferrara:1996dd,Ferrara:1996um} and cannot be varied.  The remaining 20 scalars parametrise the moduli space of Type IIB strings on 
$\AdS_3\times \Sphere^3\times\Torus^4$, which locally takes the form of a coset
\begin{equation}
  \mathcal{K}=\frac{\grpSO(5,4)}{\grpSO(5)\times \grpSO(4)}.
\end{equation}
The U-duality group $\grpSO(5,5;\Integers)$~\cite{Hull:1994ys} acts both on $\mathcal{K}$ and on the background charges. A subgroup of the U-duality group, denoted by $\mathcal{H}_{q}$ in~\cite{Larsen:1999uk}, leaves the background charges invariant, yet acts non-trivially on $\mathcal{K}$ through global identifications. As a result~\cite{Larsen:1999uk}, globally the moduli space is $\mathcal{H}_q\backslash \mathcal{K}$.

\subsection{Moduli and fixed scalars in the D1/D5 geometry}
\label{sec:summary-d1-d5-moduli}

The near-horizon limit of the D1/D5 geometry is $\AdS_3\times \Sphere^3\times\Torus^4$ supported by RR three-form flux, and the 20 moduli are:\footnote{%
  More precise definitions of the moduli will be given in section~\ref{sec:d1-d5-moduli}.%
}
\begin{enumerate}[label=\emph{(\roman*)}]
\item 9 geometric moduli of $\Torus^4$ excluding  the overall volume $v$, 
\item 6 moduli from the RR two-form potential on $\Torus^4$, 
\item 3 moduli from the self-dual part of the NSNS two-form $B^+$ on $\Torus^4$, 
\item the string coupling constant $g_s$, and
\item 1 linear combination of RR scalar and four-form potential on $\Torus^4$.\footnote{%
    We will find it convenient to treat the RR scalar $C_0$ as the modulus and the four form $C_4$ on $Torus^4$ as fixed. A more accurate description would be to say that one linear combination of these modes remains massless, and thus is a proper modulus, while the other combination obtains a mass and hence gets a fixed expectation value.%
  }
\end{enumerate}
In this background the five fixed scalars are 
\begin{enumerate}[label=\emph{(\alph*)}]
\item  3 scalars from the anti-self-dual part of the NSNS two-form potential on $\Torus^4$, 
\item the volume of $\Torus^4$, $v$
  , and 
\item a second linear combination of the RR scalar and four-form potentials.
\end{enumerate}
The values that these five scalars take varies with the moduli, and can be determined by minimising the BPS mass-charge formula for the background, which in the case of the D1/D5 system reduces to solving equations (10)-(12) of~\cite{Larsen:1999uk}. For example, at the point
\begin{equation}\label{eq:orig-mod-d1-d5}
  G_{ij} = \sqrt{v} \delta_{ij}, \qquad
  C_2 = 0, \qquad
  B^+ = 0, \qquad
  C_0 = 0 ,
\end{equation}
where $i,j=6,7,8,9$, we have
\begin{equation}\label{eq:scalars-at-origin-d1-d5}
  v = \frac{N_1}{N_5}, \qquad
  B^- = 0, \qquad
  C_4 = 0 ,
\end{equation}
where $N_1$ and $N_5$ denotes the number of D1 and D5 branes, respectively.

\subsection{Moduli and fixed scalars in the F1/NS5 geometry}
\label{sec:summary-f1-ns5-moduli}

Similarly, the near-horizon limit of the F1/NS5 geometry is $\AdS_3\times \Sphere^3\times\Torus^4$ supported by NSNS three-form flux, and the 20 moduli are: 
\begin{enumerate}[label=\emph{(\roman*)}]
\item 9 geometric moduli of $\Torus^4$ excluding  the overall volume $v$, 
\item 6 moduli from the NSNS two-form potential on $\Torus^4$, 
\item 3 moduli from the self-dual part of the RR two-form potential on $\Torus^4$, 
\item the string coupling constant $g_s$, and
\item 1 linear combination of RR scalar and four-form potential on $\Torus^4$.
\end{enumerate}
Here the five fixed scalars are 
\begin{enumerate}[label=\emph{(\alph*)}]
\item  3 scalars from the anti-self-dual part of the RR two-form potential on $\Torus^4$, 
\item the volume of $\Torus^4$, $v$, and 
\item a second linear combination of the RR scalar and four-form potentials.
\end{enumerate}
The values of the fixed scalars are determined by solving equations (15)-(17) of~\cite{Larsen:1999uk}. For example, at the point
\begin{equation}\label{eq:orig-mod-f1-ns5}
  G_{ij} = \sqrt{v} \delta_{ij}, \qquad
  B = 0, \qquad
  C^+ = 0, \qquad
  C_0 = 0
\end{equation}
we have
\begin{equation}\label{eq:scalars-at-origin-f1-ns5}
  v = \frac{N_{\text{F1}}}{k}g_s^2, \qquad
  C^- = 0, \qquad
  C_4 = 0 ,
\end{equation}
where $k$ and $N_{\text{F1}}$ are the number of NS5-branes and fundamental strings, respectively.

\section{Closed strings and moduli in D1/D5 background}
\label{sec:d1-d5-moduli}

As reviewed in section~\ref{sec:summary-d1-d5-moduli}, the near-horizon D1/D5 geometry has 20 moduli. In this section, we establish what effect these moduli have on energies of $\mathcal{H}_{(0,0)}$ states. We find that the spectrum is independent of 16 of them, and determine the influence of the remaining 4 at weak string coupling. It turns out that varying these 4 moduli has a remarkably simple effect on $\mathcal{H}_{(0,0)}$. The energy spectrum is determined in terms of the radius of curvature $R$, which depends on the 4 moduli, as well as on $N_5$.

Below we investigate how moduli enter the GS action. We find that for 16 moduli the action remains invariant up to simple field redefinitions. Of the remaining 4, one is $g_s$ whose only effect is to change $R$, via equation~\eqref{eq:rad-d1d5-bmod}. As we show below, the other 3 moduli have a non-trivial effect on the action. Luckily, it turns out that this new action is related to the original one through T-duality. As a result, the  
$\mathcal{H}_{(0,0)}$ spectrum at any value of these 3 moduli is determined from the one found in~\cite{Borsato:2016kbm,Borsato:2016xns} together with a suitable identification of $R$, which will now depend on these moduli.

\subsection{The inconsequential moduli}

The energies of states in $\mathcal{H}_{(0,0)}$ do not depend on the 16 moduli \emph{(i)}, \emph{(ii)} and \emph{(v)} of section~\ref{sec:summary-d1-d5-moduli}. It is straightforward to see that the geometric moduli of $\Torus^4$ have no effect on $\mathcal{H}_{(0,0)}$. This is because 
given any flat metric on $\Torus^4$ we can always redefine the $x^i$ to reduce to the metric~\eqref{eq:orig-mod-d1-d5} used in~\cite{Borsato:2014hja}. While such a redefinition changes the periodicity conditions of the $x^i$, this has no effect on the zero-winding zero-momentum states of $\mathcal{H}_{(0,0)}$.\footnote{\label{ftn:vol}Note that the spectrum is also independent of $v$, since we can rescale the 
$\Torus^4$ coordinates $x^i\rightarrow v^{-1/4} x^i$. The resulting change in $x^i$ periodicity conditions is also inconsequential for $\mathcal{H}_{(0,0)}$ strings.}

Next consider turning on a non-zero constant $C_2$ on $\Torus^4$. The gauge-invariant RR field strengths are defined as
\begin{equation}\label{eq:gauge-inv-rr-fs}
F_{p+1}= dC_p - C_{p-2} \wedge H,\qquad H=dB,
\end{equation}
with $H\equiv 0$ in this geometry. We see immediately that a constant $C_2$ has no effect on the field strengths, which in turn leaves the equations of motion and Bianchi identities unchanged, since these depend only on the field strengths rather than potentials (see equation~\eqref{eq:eoms-and-bianchis}). We may therefore conclude that the $\AdS_3\times \Sphere^3\times\Torus^4$ geometry with RR flux can be deformed by turning on a constant $C_2$. What is more, the background charges   do not change as we vary $C_2$.\footnote{This is consistent with the fact that a constant RR potential does not induce any additional brane charges in the D1 or D5 brane worldvolume theories.} We can see this explicitly by integrating the equations of motion/Bianchi identities for the fluxes~\cite{Marolf:2000cb}. For example, D5-brane and D3-brane charges are given by 
\begin{equation}
  \begin{aligned}
    Q_{\text{D5}} &= \frac{1}{2\kappa_0^2} \int_{\Sphere^3}F_3 = \mu_5 N_5,
    \\
    Q_{\text{D3}} &= \frac{1}{2\kappa_0^2} \int_{\Sphere^3\times\Torus^2_{ij}} \!\!\!\! F_5 + C_2 \wedge H = 0,
  \end{aligned}
\end{equation}
neither of which depends on $C_2$.\footnote{In the case of the D3-brane charge this is because both $F_5$ and $H$ are identically zero.} Similarly, one can check that the D1-, NS5- and F1-charges remain respectively $N_1$, $0$ and $0$ for all values of $C_2$. This shows that turning on a constant $C_2$ in this geometry corresponds to adjusting the values of the 6 moduli listed in \emph{(ii)} of section~\ref{sec:summary-d1-d5-moduli}. The GS action depends only on field strengths and not on potentials, and so does not change as we vary $C_2$. We conclude that the 6 $C_2$ have no effect on the energies of states in $\mathcal{H}_{(0,0)}$.

By an almost identical analysis to the above, one finds that the modulus $C_0$ also has no effect on the $\mathcal{H}_{(0,0)}$ spectrum. Turning on constant values for $C_0$ and $C_4$ leaves the field strengths unchanged and so equations of motions and Bianchi identities continue to be satisfied. Adding constant $C_0$ and $C_4$ potentials does change the F1 charge
\begin{equation}
Q_{\text{F1}} = \frac{1}{2\kappa_0^2} \int_{\Sphere^3\times\Torus^4} \!\!\!\! e^{-2\Phi} * H + C_0 F_7 + C_4 \wedge F_3 - \tfrac{1}{2} B \wedge C_2 \wedge F_3 .
\end{equation}
However, the value of $C_4$ is not an independent modulus. Rather it is determined in terms of $C_0$, $B^+$ and $C_2$ by equation~(11) of~\cite{Larsen:1999uk}. This ensures that the F1 charge is zero\footnote{While a constant $C_0$ (respectively $C_4$) induces F1 charge on the D1 (D5) worldvolume, a suitable linear combination of $C_0$ and $C_4$ has a trivial net effect.}
\begin{equation}
  Q_{\text{F1}}= \frac{1}{2\kappa_0^2} \int_{\Sphere^3\times\Torus^4} \!\!\!\! e^{-2\Phi} * H =0 .
\end{equation}
As above, since the Green-Schwarz action depends only on field strengths rather than potentials we conclude that $C_0$ has no effect on the $\mathcal{H}_{(0,0)}$ spectrum.

\subsection{The consequential moduli}

In this subsection we show that the energies of states in $\mathcal{H}_{(0,0)}$ depend on $g_s$ and $B^+$ listed as \emph{(iii)} and \emph{(iv)} in section~\ref{sec:summary-d1-d5-moduli}. The effect of varying $g_s$ is well understood. The radius of $\AdS_3$ and $g_s$ are related by equation~\eqref{eq:rad-d1d5-bmod}.
For $g_s\ll 1$, where perturbative string theory is valid, varying $g_s$ changes the energies of worldsheet excitations through the function $h(R)$ as described in equation~\eqref{eq:rr-disp-rel}.

At first sight, one might conclude that a constant $B^+$ should be just as inconsequential as a constant $C_2$. It is certainly true that adding a constant $B^+$ does not change the field strengths and so gives a consistent family of $\AdS_3$ backgrounds. However, this deformation does not correspond to turning on a modulus in the near-horizon D1/D5 geometry because the background has a non-zero D3-brane charge\footnote{A constant B field induces D3-brane charge in the D5-brane worldvolume theory through its WZ couplings~\cite{Douglas:1995bn}. As a result, we may anticipate that the $B^+$ modulus in this background will be more subtle than $C_2$.} 
\begin{equation}
  Q_{\text{D3}} =  \frac{1}{2\kappa_0^2} \int_{\Sphere^3\times\Torus^2_{ij}}F_5 - B^+\wedge F_3 \neq 0 . 
\end{equation}
To find a background with $Q_{\text{D3}}=0$ we need to further add a non-constant\footnote{The resulting $F_5\neq 0$ is self-dual and satisfies its equation of motion.} $C_4$
\begin{equation}
  dC_4=B^+\wedge F_3.
\end{equation}
The explicit expression for the geometry and fluxes is given in Appendix~\ref{sec:general-ads3-s3-t4-bg}, with the
$\AdS_3$ radius of curvature $R$ given by
\begin{equation}
  R^2=\alpha' e^\phi N_5\sqrt{1+\tfrac{1}{2}(B^+)^2}.
\label{eq:rad-curv-w-moduli}
\end{equation}
This geometry corresponds to the near-horizon limit of the fully back-reacted D1/D5 system with non-zero  $B^+$ moduli~\cite{Maldacena:1999mh,Dhar:1999ax}, which we discuss in section~\ref{sec:backreacted-d1-d5}. 

To recapitulate, turning on $B^+$ moduli in the $\AdS_3$ geometry with RR three-form flux induces a non-trivial $F_5$ flux. This clearly modifies the GS action and so should have a non-trivial effect on the energies of $\mathcal{H}_{(0,0)}$ states. Fortunately, it is possible to establish how the spectrum varies with the $B^+$ moduli without having to repeat the lengthy calculations of~\cite{Borsato:2014hja}. This is because, as is well-known~\cite{Maldacena:1999mh,Dhar:1999ax}, turning on a non-zero $B^+$ modulus is equivalent to performing T-dualities and field redefinitions of the $\Torus^4$ coordinates together with a compensating shift in the B field. As we show below, understanding the effect of these manipulations on the spectrum is straightforward. 

\subsection{TrT and String Sigma model}

Let us briefly review the procedure of~\cite{Maldacena:1999mh,Dhar:1999ax}.
In the undeformed D1/D5 background \footnote{%
  See appendix~\ref{sec:backreacted-d1-d5} for a presentation of the D1/D5 system in our conventions.%
}
perform a T-duality along one of the $\Torus^4$ directions, say $x^6$, followed by a redefinition of the $(x^6,x^7)$ variables 
\begin{equation}
  x^6\rightarrow x^6\cos\varphi+x^7\sin\varphi,
  \qquad
  x^7\rightarrow x^7\cos\varphi-x^6\sin\varphi,
  \label{eqn:x67-redef}
\end{equation}
and then a T-duality back along $x^6$. We will call this a TrT transformation, since it is similar to, though distinct from, the TsT transformations, much used in $\AdS_5$~\cite{Lunin:2005jy,Frolov:2005dj}. Starting with a square $\Torus^2$ of area $A$ with no B field, a TrT transformation maps to a square $\Torus^2$ of area $\tilde{A}=A(\cos^2\varphi+A^2\sin^2\varphi)^{-1/2}$ and a B-field $\tilde{B}_{67}=(1-A^2)(\cos^2\varphi+A^2\sin^2\varphi)^{-1/2}$. We can also perform a TrT transformation, parametrised by $\psi$, in the $(x^8,x^9)$ directions. 
respectively. 
The resulting TrT${}^2$ backgrounds carry D1-, D3- and D5-brane charges.\footnote{Explicit expressions for the metric and other fields in are presented in section~\ref{sec:backreacted-d1-d5-b}.}
However, we can further add a constant B-field 
\begin{equation}
  \tilde{B}_{67} \rightarrow \tilde{B}_{67} + b_{67} , \qquad
  \tilde{B}_{89} \rightarrow \tilde{B}_{89} + b_{89} ,
  \label{eq:extra-cst-b}
\end{equation}
which, if chosen judiciously~\cite{Maldacena:1999mh,Dhar:1999ax}, (see equation~\eqref{eq:b67-b89-no-D3}) precisely cancels the D3-brane charges of the TrT${}^2$ background. Restricting to
\begin{equation}
  \varphi=\psi,
  \label{eq:trt-params-rel}
\end{equation}
sets $\Vol(T^2_{67}) = \Vol(T^2_{89})$. In summary,  turning on the $B^+_{67}$ modulus is equivalent to a TrT${}^2$ deformation with $\varphi=\psi$ and an additional constant B field. 

Let us now consider the effect of these manipulations on the worldsheet action.
A constant shift of the B field~\eqref{eq:extra-cst-b}, has no effect on the GS action since this depends only on field strengths. While such a B field does modify the periodicity conditions of the $\Torus^4$ coordinates $x^i$, this has no effect on the zero-momentum, zero-winding strings that we are interested in. As a result, the  $\mathcal{H}_{(0,0)}$ spectrum of the theory deformed by a $B^+$ modulus is the same as the $\mathcal{H}_{(0,0)}$ spectrum of the TrT$^2$ theory with $\psi=\varphi$. In turn,  the $\mathcal{H}_{(0,0)}$ spectrum of the TrT$^2$ theory can be obtained from the original undeformed theory by analysing the effects of T-dualities and field redefinitions, as we now describe.

The $\Torus^4$ coordinates $x^i$ enter the GS action only through their derivatives, since they correspond to $U(1)$ isometries
\begin{equation}
  \label{eq:gen-string-action}
  \begin{aligned}
    S&=-\frac{1}{4\pi}\int d^2\sigma \Bigl(
    \gamma^{\alpha\beta}\partial_\alpha x^i\partial_\beta x^j G_{ij}
    -\epsilon^{\alpha\beta}\partial_\alpha x^i\partial_\beta x^j B_{ij}
    \\ &\qquad\qquad\qquad\qquad
    +2\partial_\alpha x^i
    \bigl(\gamma^{\alpha\beta}U_{\beta,i}-\epsilon^{\alpha\beta}V_{\beta,i}\bigr)+\mathcal{L}_{\text{rest}}\Bigr).
  \end{aligned}
\end{equation}
The explicit couplings $G, B,V$ and $U$ can be found in~\cite{Borsato:2014hja} to quadratic order in fermions.\footnote{As a result, the fermions are automatically uncharged under the $U(1)$ isometries of $T^4$, unlike what happens in TsT backgrounds~\cite{Alday:2005ww}.} The Noether currents for the $x^i$ shift symmetries are 
\begin{equation}
  J^\alpha_i(x)=-\left(\gamma^{\alpha\beta}\partial_\beta x^j G_{ij}
    -\epsilon^{\alpha\beta}\partial_\beta x^j B_{ij}
    +\gamma^{\alpha\beta}U_{\beta,i}-\epsilon^{\alpha\beta}V_{\beta,i}\right).
  \label{eq:-noeth-curr-gen}
\end{equation}
The TrT-transformed action takes the same general form as~\eqref{eq:gen-string-action}, for the dual coordinates $\tilde{x}^i$ with couplings $\tilde{G}$, $\tilde{B}$, $\tilde{U}$ and $\tilde{V}$.
\footnote{Explicit expressions for $\tilde{G}$, $\tilde{B}$, $\tilde{U}$ and $\tilde{V}$ can be found in appendix~\ref{sec:trt-transformed-GS}.}
The corresponding Noether currents are
\begin{equation}
  \tilde{J}^\alpha_i(\tilde{x})=-\left(\gamma^{\alpha\beta}\partial_\beta \tilde{x}^j \tilde{G}_{ij}
    -\epsilon^{\alpha\beta}\partial_\beta \tilde{x}^j \tilde{B}_{ij}
    +\gamma^{\alpha\beta}\tilde{U}_{\beta,i}
    -\epsilon^{\alpha\beta}\tilde{V}_{\beta,i}\right),
  \label{eq:-noeth-curr-gen-TrT}
\end{equation}
and the coordinates $x^{6,7}$ and $\tilde{x}^{6,7}$ are related via
\begin{equation}
  \label{eq:rel-x-xtilde}
  \begin{aligned}
    \partial_\alpha \tilde{x}^6 &= \cos\varphi \, \partial_\alpha x^6 
    -\sin\varphi \left( \epsilon_{\alpha\beta}\gamma^{\beta\delta}\partial_\delta x^i G_{i7}-\partial_\alpha x^i B_{i7}
      -\epsilon_{\alpha\beta}\gamma^{\beta\delta}U_{\delta, 7}-V_{\alpha, 7}\right),
    \\
    \partial_\alpha \tilde{x}^7 &= \cos\varphi \, \partial_\alpha x^7 
    +\sin\varphi \left( \epsilon_{\alpha\beta}\gamma^{\beta\delta}\partial_\delta x^i G_{i6}-\partial_\alpha x^i B_{i6}
      -\epsilon_{\alpha\beta}\gamma^{\beta\delta}U_{\delta, 6}-V_{\alpha, 6}\right).    
  \end{aligned}
\end{equation}
As a result, the currents $J^\alpha_i$ and $\tilde{J}^\alpha_i$ are B\"acklund transforms of each other
\begin{equation}
  \tilde{J}^\alpha_6 = \cos\varphi \, J^\alpha_6 - \sin\varphi \, \epsilon^{\alpha\beta}\partial_\beta x^7,
  \qquad
  \tilde{J}^\alpha_7 = \cos\varphi \, J^\alpha_7 + \sin\varphi \, \epsilon^{\alpha\beta}\partial_\beta x^6,
\end{equation}
and we have
\begin{equation}
  \partial_\sigma \tilde{x}^6 = \cos\varphi \, \partial_\sigma x^6 - \sin\varphi \, J^\tau_7, \qquad
  \partial_\sigma \tilde{x}^7 = \cos\varphi \, \partial_\sigma x^7 + \sin\varphi \, J^\tau_6.
\end{equation}
Integrating the above with respect to $\sigma$ and reasoning in a similar way to~\cite{Frolov:2005dj}, we find that instead of analysing states with conventional periodicity on a TrT-transformed background
\begin{equation}
\tilde{x}^6(2\pi)-\tilde{x}^6(0)=\tilde{w}_6,\qquad
\tilde{x}^7(2\pi)-\tilde{x}^7(0)=\tilde{w}_7,
\end{equation}
we can consider states in the \emph{untransformed} background with twisted periodicity conditions for $x^{6,7}$
\begin{equation}
  x^6(2\pi)-x^6(0)=\sec\varphi \tilde{w}_6+P_7\tan\varphi ,
  \qquad
  x^7(2\pi)-x^7(0)=\sec\varphi \tilde{w}_7-P_6\tan\varphi .
\label{eq:tw-per-cond}
\end{equation}
Above, $P_{6,7}$ are  charges (momenta) of the currents $J^\alpha_{6,7}$.  What is more, since $J^\tau_i$ is the momentum variable conjugate to $x^i$, we can in fact relate the TrT-transfomed variables to the original ones through a canonical transformation
\begin{equation}
  \begin{aligned}
    \tilde{p}_6 &= \cos\varphi \, p_6 - \sin\varphi \, x_7',
    &
    \tilde{x}^\prime_6 &= \cos\varphi \, x^\prime_6 - \sin\varphi \, p_7,
    \\
    \tilde{p}_7 &= \cos\varphi \, p_7 + \sin\varphi \, x_6',
    &
    \tilde{x}^\prime_7&=\cos\varphi \, x^\prime_7 + \sin\varphi \, p_6.
  \end{aligned}
\end{equation}
We therefore conclude that the spectrum of closed strings in a TrT-transformed background will be the same as the spectrum of strings with twisted periodicity conditions~\eqref{eq:tw-per-cond} in the original background. This is true for \emph{all} perturbative closed string states, since the TrT transformation is an exact T-duality symmetry of the theory. 

In the $\mathcal{H}_{(0,0)}$ sector of most interest in this paper, states have zero winding and momentum, and it is easy to see that equation~\eqref{eq:tw-per-cond} maps periodic states to periodic states. As a result, the energies of $\mathcal{H}_{(0,0)}$ states before and after TrT transformations are the same. At weak coupling, they depend only on the respective radii of curvature of the two theories, which are equal to one another, and taking into account the shift of the dilaton under T-duality can be written as
\begin{equation}
R^2 = \alpha' e^\phi N_5
= \alpha' e^{\tilde{\phi}} \left(N_5 \cos^2\varphi+N_1 \sin^2\varphi\right) .
\end{equation}
Note that this expression is written in terms of the original D-brane charges $N_1$ and $N_5$. This expression gets further corrected by the constant shift of the B field, as discussed in section~\ref{sec:backreacted-d1-d5-b}.
From the arguments given above equation~\eqref{eq:gen-string-action}, we then find that the energies of $\mathcal{H}_{(0,0)}$ states depend on the value of the $B^+$ moduli through the dependence of $R$ in equation~\eqref{eq:rad-curv-w-moduli}.

In~\cite{Borsato:2016kbm,Borsato:2016xns} the energies of $\mathcal{H}_{(0,0)}$ strings  were found as solutions of (essentially algebraic) BEs. The explicit calculations were carried out at the point in moduli space given in equation~\eqref{eq:orig-mod-d1-d5} and at small 
$g_s$. It was shown that the 2-to-2 worldsheet S matrix is fixed by symmetries alone~\cite{Borsato:2012ud,Borsato:2013qpa,Borsato:2014exa,Borsato:2014hja}.\footnote{More precisely symmetries fix the matrix part of the S matrix, leaving undetermined overall normalisations known as dressing factors. The dressing factors are then determined by solving suitable crossing equations~\cite{Borsato:2013hoa,Borsato:2016xns}.} and satisfies the Yang-Baxter equation. This determines the complete worldsheet scattering and hence the spectrum through the Bethe Ansatz. 

In principle, this concludes our analysis of the dependence of the $\mathcal{H}_{(0,0)}$ spectrum on $g_s$ and $B^+$. In the remainder of this section, we show more explicitly that deriving the 2-to-2 S matrix following~\cite{Borsato:2014hja}, is fully compatible with the TrT deformations we have discussed above. We pay particular attention to gauge-fixing, expressions for supercharges and the off-shell algebra $\mathcal{A}$ under TrT. The reader not interested in these technical details may wish to proceed directly to the next section.

T-duality along $\Torus^4$ directions can be used to map the background considered here to other Type IIB backgrounds: the D3/D3' background or the D1/D5 background with $N_1$ and $N_5$ swapped, or related Type IIA backgrounds. The energies of $\mathcal{H}_{(0,0)}$ states will remain unchanged under such T-dualities and since the states carry no winding or momentum T-duality will map them to $\mathcal{H}_{(0,0)}$ states in the dual background.

\subsection{TrT and gauge fixing}
\label{sec:trt-eoms}

The string two-body S matrix found in~\cite{Borsato:2014hja} was determined in a particular gauge. It is therefore worth checking that the gauge-fixing is compatible with TrT transformations. The gauge fixing is a two-step process: fixing kappa gauge, and 
fixing uniform light-cone gauge. 

In~\cite{Borsato:2014hja} a kappa-gauge that is particularly well adapted to the underlying integrability was used. This kappa-gauge is a simple projection on the (spectator) fermions, and as result it commutes with T-duality and redefinitions of the $\Torus^4$ bosons. We can therefore apply TrT transformations directly at the level of the kappa-gauge-fixed action given in equations (2.60) and (G.6) of~\cite{Borsato:2014hja}.

Before proceeding to uniform light-cone gauge, we briefly comment on the relations between the equations of motion in theories related by TrT transformations.The $x^i$ and $\tilde{x}^i$ equations of motion are given by  current conservation equations
\begin{equation}
  \partial _\alpha J^\alpha = 0, \qquad
  \partial _\alpha \tilde{J}^\alpha = 0,
\end{equation}
and are equivalent to one another when equation~\eqref{eq:rel-x-xtilde} is used. The equations of motion for the other fields, including the Virasoro constraints for the worldsheet metric $\gamma^{\alpha\beta}$, remain the same in the original and TrT-transformed background upon using the relations between $G,B,U,V$ and $\tilde{G},\tilde{B},\tilde{U},\tilde{V}$, as well as equation~\eqref{eq:rel-x-xtilde}. To see this, let us introduce a collective field 
\begin{equation}
  \omega(\tau,\sigma)=\left\{
    x^\pm,y^i,z^i,\chi,\eta,
    \partial_\alpha x^\pm, \partial_\alpha y^i, \partial_\alpha z^i,\partial_\alpha\chi,\partial_\alpha\eta
  \right\}.
\end{equation}
Then one can show that
\begin{multline}
  \label{eq:equiv-eoms}
  \gamma^{\alpha\beta}\partial_\alpha x^i\partial_\beta x^j \frac{\partial G_{ij}}{\partial\omega}
  -\epsilon^{\alpha\beta}\partial_\alpha x^i\partial_\beta x^j \frac{\partial B_{ij}}{\partial\omega}
  +2\partial_\alpha x^i
  \left(\gamma^{\alpha\beta}\frac{\partial U_{\beta,i}}{\partial\omega}
    -\epsilon^{\alpha\beta}\frac{\partial V_{\beta,i}}{\partial\omega}\right)+\frac{\mathcal{L}_{\text{rest}}}{\partial\omega} =
  \\
  \gamma^{\alpha\beta}\partial_\alpha \tilde{x}^i\partial_\beta \tilde{x}^j \frac{\partial \tilde{G}_{ij}}{\partial\omega}
  -\epsilon^{\alpha\beta}\partial_\alpha \tilde{x}^i\partial_\beta \tilde{x}^j \frac{\partial \tilde{B}_{ij}}{\partial\omega}
  +2\partial_\alpha \tilde{x}^i
  \left(\gamma^{\alpha\beta}\frac{\partial \tilde{U}_{\beta,i}}{\partial\omega}
    -\epsilon^{\alpha\beta}\frac{\partial \tilde{V}_{\beta,i}}{\partial\omega}\right)+\frac{\mathcal{\tilde{L}}_{\text{rest}}}{\partial\omega}
  .
\end{multline}
Hence the equations of motion for the non-$\Torus^4$ sigma-model fields are the same before and after a TrT transformation. Similarly, one can show that
\begin{equation}
  \frac{\delta S}{\delta \gamma^{\alpha\beta}}
  =
  \frac{\delta \tilde{S}}{\delta \gamma^{\alpha\beta}},
\end{equation}
confirming that the Virasoro constraints of the two theories are also the same.

In uniform light-cone gauge~\cite{Arutyunov:2005hd,Borsato:2014hja} we set $x^+\equiv \partial_\alpha(\phi-t)/2=\tau$ and its conjugate momentum, $p_-=1$. The Virasoro constraints then determine the non-dynamical field 
$x^-\equiv \partial_\alpha(\phi-t)/2$ in terms of the transverse excitations and the worldsheet metric $\gamma^{\alpha\beta}$, 
while the $x^\pm$ equations of motion are used to fix $\gamma^{\alpha\beta}$. As we have shown above the $x^\pm$ and $\gamma^{\alpha\beta}$ equations of motion are invariant under TrT transformations. As a result, gauge-fixing the TrT-transformed action gives the same expressions for $\dot{x}^-$, $\pri{x}^-$ and $\gamma^{\alpha\beta}$ as those found in the original background~\cite{Borsato:2014hja}; the dependence on $\partial_\alpha x^i$ can be re-expressed in terms of $\partial_\alpha \tilde{x}^i$ by using equation~\eqref{eq:rel-x-xtilde}, as well as the relations between $G,B,U,V$ and $\tilde{G},\tilde{B},\tilde{U},\tilde{V}$. We conclude that gauge-fixing the GS action is compatible with TrT transformations.

\subsection{Supercharges}

In~\cite{Borsato:2014hja}, supercurrents $Q^\alpha$ were constructed, in terms of  transverse fields and their derivatives, as well as the ubiquitous non-local prefactor $e^{\pm i x^-}$. The conservation of these supercurrents
was checked using the equations of motion. In these expressions for the supercurrents
the torus bosons $x^i$ enter only through the first derivatives $\partial_\alpha x^i$.

Therefore, in order to find the supercharges in the TrT theory, one can use the inverse of equation~\eqref{eq:rel-x-xtilde} to re-express $Q^\alpha$ in terms of the transverse fields of the TrT-transformed theory, as well as the pre-factor $e^{\pm i x^-}$. As we reviewed in the previous sub-section, $x^-$ is determined via the Virasoro constraints, which are the same in the TrT-related theories upon using the relations between $G,B,U,V$ and $\tilde{G},\tilde{B},\tilde{U},\tilde{V}$, as well as equation~\eqref{eq:rel-x-xtilde}. To summarise, we can express the supercurrents $Q^\alpha$, exclusively in terms of the fields that enter the TrT transformed action 
$\tilde{S}$. We will denote by $\tilde{Q}^\alpha$ this putative supercurrent.

It remains to be checked whether  $\tilde{Q}^\alpha$ is conserved when equations of motion  derived from $\tilde{S}$ are used. This however, has to be, because: (i) the equations of motion for all fields other than $\tilde{x}^i$ (including $\gamma^{\alpha\beta}$) are
the same in the two theories; (ii) the $\tilde{x}^i$ equations of motion together with equation~\eqref{eq:rel-x-xtilde} are equivalent to the $x^i$ equations of motion. Therefore,  $\tilde{Q}^\alpha$ is conserved upon using equations of motion derived from $\tilde{S}$.

\subsection{Determining the off-shell algebra \texorpdfstring{$\mathcal{A}$}{A}}

Finally,  we turn to the algebra $\mathcal{A}$ of supercharges which commute with the Hamiltonian. In order to find the commutation relations of $\mathcal{A}$ in the undeformed theory, it was necessary to redefine the fermions in order to obtain a canonical kinetic term for the fermions; see Appendix I of~\cite{Borsato:2014hja} for details. In principal, an analogous computation should be performed in the TrT transformed background. Explicitly one needs to find
\begin{equation}
  \frac{\delta^2 \tilde{S}}{\delta \bar{\eta}\delta \dot{\eta}},
  \qquad
  \frac{\delta^2 \tilde{S}}{\delta \bar{\chi}\delta \dot{\chi}},
  \qquad
  \frac{\delta^2 \tilde{S}}{\delta \bar{\chi}\delta \dot{\eta}},
  \qquad
  \frac{\delta^2 \tilde{S}}{\delta \bar{\eta}\delta \dot{\chi}}.
\end{equation}
It is easy to see that the first two expressions involve only $\tilde{\mathcal{L}}_{\text{rest}}$. However, to quadratic order in fermions $\tilde{\mathcal{L}}_{\text{rest}}=\mathcal{L}_{\text{rest}}$, hence equations (I.1) and (I.2) of~\cite{Borsato:2014hja} remain the same after TrT transformations.
On the other hand the mixed $\eta$-$\chi$ terms at first appear to change after a TrT transformation. We note that these terms do not involve $\tilde{\mathcal{L}}_{\text{rest}}$, and to the order that we are working to
\begin{equation}
  \frac{\partial^2 \tilde{G}}{\partial \bar{\chi}\partial \dot{\eta}}
  =\frac{\partial^2 \tilde{B}}{\partial \bar{\chi}\partial \dot{\eta}}=0.
\end{equation}
As a result, the only non-zero contributions to $\eta$-$\chi$ couplings come from $\tilde{U}$ and $\tilde{V}$ terms, and to the order we are working these 
give
\begin{equation}
  \begin{aligned}
    -2\pi\frac{\delta^2 \tilde{S}}{\delta \bar{\chi}\delta \dot{\eta}}
    &=
    \partial_\alpha{x}^i\gamma^{\alpha\beta}
    \frac{\partial^2 \tilde{U}_{\beta,i}}{\partial \bar{\chi}\partial \dot{\eta}}
    -\partial_\alpha{x}^i\epsilon^{\alpha\beta}
    \frac{\partial^2 \tilde{V}_{\beta,i}}{\partial \bar{\chi}\partial \dot{\eta}}
    ,
    \\
    -2\pi\frac{\delta^2 \tilde{S}}{\delta \bar{\eta}\delta \dot{\chi}}
    &=
    \partial_\alpha{x}^i\gamma^{\alpha\beta}
    \frac{\partial^2 \tilde{U}^{(f)}_{\beta,i}}{\partial \bar{\eta}\partial \dot{\chi}}
    -\partial_\alpha{x}^i\epsilon^{\alpha\beta}
    \frac{\partial^2 \tilde{V}^{(f)}_{\beta,i}}{\partial \bar{\eta}\partial \dot{\chi}}
    .
  \end{aligned}
\end{equation}
Given the trivial contributions of $\tilde{G}$, $\tilde{B}$ to the mixed terms, and the fact that $\gamma^{\alpha\beta}$ does not depend on fermions, at quadratic order in fermions the above expressions can be re-written as
\begin{equation}
  \begin{aligned}
    -2\pi\frac{\delta^2 \tilde{S}}{\delta \bar{\chi}\delta \dot{\eta}}
    &=
    \frac{\partial}{\partial \dot{\eta}}\biggl[
    \gamma^{\alpha\beta}\partial_\alpha \tilde{x}^i\partial_\beta \tilde{x}^j \frac{\partial \tilde{G}_{ij}}{\partial\bar{\chi}}
    -\epsilon^{\alpha\beta}\partial_\alpha \tilde{x}^i\partial_\beta \tilde{x}^j \frac{\partial \tilde{B}_{ij}}{\partial\bar{\chi}}
    \\ 
    &\qquad\qquad
    +2\partial_\alpha \tilde{x}^i
    \left(\gamma^{\alpha\beta}\frac{\partial \tilde{U}_{\beta,i}}{\partial\bar{\chi}}
      -\epsilon^{\alpha\beta}\frac{\partial \tilde{V}_{\beta,i}}{\partial\bar{\chi}}\right)+\frac{\tilde{\mathcal{L}}_{\text{rest}}}{\partial\bar{\chi}}
    \biggr] ,
    \\
    -2\pi\frac{\delta^2 \tilde{S}}{\delta \bar{\eta}\delta \dot{\chi}}
    &=
    \frac{\partial}{\partial \dot{\chi}} \biggl[
    \gamma^{\alpha\beta}\partial_\alpha \tilde{x}^i\partial_\beta \tilde{x}^j \frac{\partial \tilde{G}_{ij}}{\partial\bar{\eta}}
    -\epsilon^{\alpha\beta}\partial_\alpha \tilde{x}^i\partial_\beta \tilde{x}^j \frac{\partial \tilde{B}_{ij}}{\partial\bar{\eta}}
    \\ 
    &\qquad\qquad
    +2\partial_\alpha \tilde{x}^i
    \left(\gamma^{\alpha\beta}\frac{\partial \tilde{U}_{\beta,i}}{\partial\bar{\eta}}
      -\epsilon^{\alpha\beta}\frac{\partial \tilde{V}_{\beta,i}}{\partial\bar{\eta}}\right)+\frac{\tilde{\mathcal{L}}_{\text{rest}}}{\partial\bar{\eta}}
    \biggr] .
  \end{aligned}
\end{equation}
We now observe that the expressions inside the square-brackets above are precisely the same as the right hand side of equation~\eqref{eq:equiv-eoms} 
with $\omega=\bar{\chi},\bar{\eta}$. Therefore, using equation~\eqref{eq:equiv-eoms} 
we conclude that
\begin{equation}
  \frac{\delta^2 \tilde{S}}{\delta \bar{\chi}\delta \dot{\eta}}=
  \frac{\delta^2  S}{\delta \bar{\chi}\delta \dot{\eta}},
  \qquad
  \frac{\delta^2 \tilde{S}}{\delta \bar{\eta}\delta \dot{\chi}}=
  \frac{\delta^2  S}{\delta \bar{\eta}\delta \dot{\chi}}.
\end{equation}
In other words, equation (I.3) of~\cite{Borsato:2014hja} remains unchanged\footnote{Recall that one uses equation~\eqref{eq:rel-x-xtilde} to swap between $x^i$ and $\tilde{x}^i$.} As a result, the redefinition of fermions in the TrT-transformed background in order to obtain a canonical kinetic term,  is the same as the one used in Appendix I of~\cite{Borsato:2014hja}. From this we finally conclude that the commutation relations for $\mathcal{A}$ are also the same as~\cite{Borsato:2014hja}.

\section{Closed strings and moduli in F1/NS5 background}
\label{sec:f1-ns5-moduli}

The near-horizon F1/NS5 geometry has 20 moduli, which we summarised in section~\ref{sec:summary-f1-ns5-moduli}. We would like to understand their effect on energies of zero-winding zero-momentum closed string states, which we will continue to denote by $\mathcal{H}_{(0,0)}$. 
Here too, we find that the spectrum is independent of 16 of them, and determine the influence of the remaining 4 at small $g_s$. As in the previous section, our analysis will primarily rely on the effect that the moduli have on the worldsheet Green-Schwarz action.

\subsection{The inconsequential moduli}

The energies of states in $\mathcal{H}_{(0,0)}$ do not depend on the 16 moduli \emph{(i)}, \emph{(ii)} and \emph{(iv)} of section~\ref{sec:summary-f1-ns5-moduli}. As in the D1/D5 background, the geometric moduli of $\Torus^4$ have no effect on $\mathcal{H}_{(0,0)}$, because they can be absorbed into suitable redefinitions of the $x^i$. The string action and periodicity conditions of $\mathcal{H}_{(0,0)}$ are also independent of $\Vol(\Torus^4)$ as before (see footnote~\ref{ftn:vol}). The string coupling constant $g_s$ and $\Vol(\Torus^4)$ are related to one another through equation (17) of~\cite{Larsen:1999uk}.  Since $g_s$ enters the GS action only through this relation to $\Vol(T^4)$, we conclude that at small string coupling $g_s$ has no effect on $\mathcal{H}_{(0,0)}$ energies.

Turning on a constant $B \neq 0$ gives a consistent background since $H$ is unchanged and the gauge-invariant RR field strengths~\eqref{eq:gauge-inv-rr-fs} remain equal to zero. One can easily check that the F1 and NS5 charges of the background are unchanged
\begin{equation}
  \begin{aligned}
    Q_{\text{NS5}} &= \frac{1}{2\kappa_0^2} \int_{\Sphere^3}H = \mu_5 k ,
    \\
    Q_{\text{F1}} &= \frac{1}{2\kappa_0^2} \int_{\Sphere^3\times\Torus^4} \!\!\! e^{-2\Phi} * H 
    + C_0 F_7 + C_4 \wedge dC_2  + \tfrac{1}{2} H \wedge C_2 \wedge C_2 = \mu_1 N_{\text{F1}} .
  \end{aligned}
\end{equation}
and that the D5-, D3- and D1-brane charges are  zero
\begin{equation}
  \begin{aligned}
    Q_{\text{D5}} &= \frac{1}{2\kappa_0^2}\int_{\Sphere^3}F_3+ C_0 H =0 ,
    \\
    Q_{\text{D3}} &= \frac{1}{2\kappa_0^2}\int_{\Sphere^3\times\Torus^2_{ij}}F_5 + C_2\wedge H = 0 ,
    \\
    Q_{\text{D1}} &= \frac{1}{2\kappa_0^2}\int_{\Sphere^3\times\Torus^4}F_7 + C_4\wedge H = 0 ,
  \end{aligned}
\end{equation}
because the RR potentials and RR gauge-invariant field strengths are all zero. In other words, turning on the $B$ moduli is accomplished in the geometry by setting $B$ to a non-zero constant on $\Torus^4$. The GS action depends only on gauge-invariant field strengths ($H$ and $F_p$) and so does not change as we vary $B$. We conclude that the 6 $B$ moduli have no effect on $\mathcal{H}_{(0,0)}$ energies.

\subsection{The consequential moduli}

In this subsection we show that the energies of states in $\mathcal{H}_{(0,0)}$ depend on $C_0$ and $C^+_2$ listed as \emph{(iii)} and \emph{(v)} in section~\ref{sec:summary-f1-ns5-moduli}. In fact, turning on a particular $C^+_2$ modulus, is equivalent to turning on $C_0$ and $C_4$, as can be seen by performing two T-dualities on $\Torus^4$. As a result, we will first focus on turning on just the $C_0$ modulus.

Let us then consider a background with constant RR potentials\footnote{%
  Note that we let the RR potential $C_0$ take arbitrary real values. Using the $\grpSL(2,\Integers)$ symmetry of type IIB string theory we can shift $C_0$ by an integer: $C_0 \to C_0 - n$. This transformation also acts on the potentials $C_2$ and $C_4$ as a shift $C_2 \to C_2 - n B$ and $C_4 \to C_4 - \tfrac{n}{2} B \wedge B$. Together, these transformations leave the gauge invariant field strengths $F_3$ and $F_5$ unchanged, but shift the D-brane charges. We therefore prefer to keep the charges fixed and not put any constraints on the value of $C_0$.%
}%
\begin{equation}
  C_0 = c_0 , \qquad
  C_4 = -c_0 \, e^6 \wedge e^7 \wedge e^8 \wedge e^9 .
\label{eq:co-mod-cst}
\end{equation}
Since $H$ is non-vanishing, this gives rise to non-zero RR three- and seven-form field strengths
\begin{equation}\label{eq:c0-mod-fs}
  \begin{aligned}
    F_3 &= dC_2-C_0 H = - c_0\, H=-c_0\,k\,\bigl( \Omega_{\AdS_3} + \Omega_{\Sphere^3} \bigr) ,
    \\
    F_7 &= dC_6-C_4 \wedge H =c_0\,k\,\bigl( \Omega_{\AdS_3} + \Omega_{\Sphere^3} \bigr) \wedge e^6 \wedge e^7 \wedge e^8 \wedge e^9.
  \end{aligned}
\end{equation}
It is straightforward to check that the D5-, D3- and D1-brane charges are all zero
\begin{equation}
  \begin{aligned}
    Q_{\text{D5}} &= \frac{1}{2\kappa_0^2} \int_{\Sphere^3} F_3+ C_0 H
    = \frac{1}{2\kappa_0^2} \int_{\Sphere^3} -C_0 H+ C_0 H 
    =0,
    \\
    Q_{\text{D3}} &= \frac{1}{2\kappa_0^2} \int_{\Sphere^3\times\Torus^2_{ij}} \!\!\!\! F_5 + C_2\wedge H = 0,
    \\
    Q_{\text{D1}} &=
    \frac{1}{2\kappa_0^2} \int_{\Sphere^3\times\Torus^4} \!\!\!\! F_7 + C_4\wedge H = 
    \frac{1}{2\kappa_0^2} \int_{\Sphere^3\times\Torus^4} \!\!\!\! -C_4\wedge H + C_4\wedge H = 0.
  \end{aligned}
\end{equation}
The NS5-brane charge remains unchanged, since $H$ stays the same, as does the F1 charge
\begin{equation}
  \begin{aligned}
    Q_{\text{F1}} &= \frac{1}{2\kappa_0^2} \int_{\Sphere^3\times\Torus^4} \!\!\! e^{-2\Phi} * H 
    + C_0 F_7 + C_4 \wedge dC_2  + \tfrac{1}{2} H \wedge C_2 \wedge C_2
    \\
    &= \frac{1}{2\kappa_0^2} \int_{\Sphere^3\times\Torus^4} \!\!\! e^{-2\Phi} * H 
    + C_0 F_7 
    =k(g_s^{-2} +c_0^2) \Vol(\Torus^4) = \mu_1 N_{\text{F1}}.
  \end{aligned}
\end{equation}
The last equality follows from equation~(17) of~\cite{Larsen:1999uk}, which determines 
$\Vol(\Torus^4)$ in terms of the moduli. In summary,  turning on a constant value of the $C_0$ modulus is implemented in the geometry not just through constant RR zero- and four-form potentials~\eqref{eq:co-mod-cst}, but also through induced three- and seven-form RR field strengths~\eqref{eq:c0-mod-fs}, with the $\AdS_3$ radius of curvature $R$ given by\footnote{This geometry corresponds to the near-horizon limit of a fully back-reacted F1/NS5 system with non-zero  $C_0$, which we obtain by U-duality from the solutions~\cite{Maldacena:1999mh,Dhar:1999ax} in section~\ref{sec:backreacted-f1-ns5-c0}.}
\begin{equation}
R^2=\alpha' k \sqrt{1+g_s^2c_0^2}.
\label{eq:rad-w-c0-mod}
\end{equation}

The non-zero RR field strengths~\eqref{eq:c0-mod-fs} have an important consequence on the GS action in this background: the world-sheet action takes the same \emph{form} as the action used to analyse mixed flux backgrounds~\cite{Hoare:2013ida,Lloyd:2014bsa}. This is because, from the point of view of the GS action, a non-zero $F_3$ generated by a non-trivial $C_2$, or by $C_0H$ are completely equivalent.\footnote{The physical interpretation of the two backgrounds is of course different. The mixed flux background investigated in~\cite{Hoare:2013ida,Lloyd:2014bsa} corresponds to the near-horizon limit of non-threshold bound states of F1/NS5- and D1/D5-branes, while the background studied in this section is a marginal deformation of the F1/NS5-brane near-horizon geometry.} As a result, the exact worldsheet S matrix found in~\cite{Hoare:2013ida,Lloyd:2014bsa} applies directly to the analysis of the $\mathcal{H}_{(0,0)}$ spectrum of the F1/NS5-brane theory deformed by the $C_0$ modulus. We simply need to relate the parameters used there to those used here! In~\cite{Hoare:2013ida,Lloyd:2014bsa} the fluxes are
\begin{equation}
e^{\phi} F_3=\tilde{q}\,\bigl( \Omega_{\AdS_3} + \Omega_{\Sphere^3} \bigr)
,\qquad
H = q\,\bigl( \Omega_{\AdS_3} + \Omega_{\Sphere^3} \bigr),
\end{equation}
together with the condition $q^2+\tilde{q}^2=1$. So, replacing 
\begin{equation}
  \label{eq:mixedf-reexp-param}
  \tilde{q}\rightarrow -g_s c_0 k \frac{\alpha'}{R^2},\qquad
  q\rightarrow k \frac{\alpha'}{R^2},
\end{equation}
where the $\AdS_3$ radius is given in equation~\eqref{eq:rad-w-c0-mod} above, we obtain the complete S matrix of the $C_0$ deformation of the pure NSNS $\AdS_3$ geometry.

The $\mathcal{H}_{(0,0)}$ spectrum consists of the BMN vacuum $\left|0\right>_{\text{BMN}}$ on top of which we can act with magnon-like creation operators denoted schematically as
\begin{equation}
\alpha^{I_1}_{p_1}{}^\dagger\dots\alpha^{I_K}_{p_K}{}^\dagger
\left|0\right>_{\text{BMN}} ,
\end{equation}
where the indices $I_1,\dotsc,I_K$ label the excitations above the BMN vacuum~\cite{Berenstein:2002jq}.
Each of the $\alpha^{I_1}_{p_1}{}^\dagger$ carries a momentum $p_i$ and has an energy
\begin{equation}
  E(p_i)=\sqrt{\left(m_i + \tfrac{k p_i}{2\pi}\right)^2+4h^2(R)\sin^2\left(\tfrac{p_i}{2}\right)},
\end{equation}
with the total energy of a state being the sum of the magnon energies. The strength of the worldsheet interactions is governed by the function $h(R)$, which in the mixed-flux backgrounds took the form
\begin{equation}
h(R)=\frac{\tilde{q}}{2\pi} \frac{R^2}{\alpha'} + \mathcal{O}(R^0).
\end{equation}
Therefore, for the background obtained by a $C_0$ marginal deformation of the pure NSNS flux theory, we have 
\begin{equation}
h(R)=-\frac{g_s c_0 k}{2\pi} + \mathcal{O}(R^0).
\end{equation}
Notice that something rather remarkable happens: the strength of the worldsheet interactions is now proportional to $c_0$, and it is this parameter that plays the analogue of the 't Hooft coupling $\lambda$ that conventionally interpolates between the weakly and strongly coupled regimes. At small $c_0$ the interactions are weak, with the dispersion relation becoming linear in the $c_0$ going to zero limit. As $c_0$ increases, the interactions become more important, modifying the dispersion relation. Throughout this range the magnon momenta $p_i$ satisfy Bethe Equations derived in~\cite{Baggio:2017kza}. The above conclusions are all valid to leading order in the large-$R$ limit, and we expect the function $h(R)$ to receive corrections when $R$ becomes small. It would be interesting to understand these in order to connect to the recent investigations of the $k=1$ theory~\cite{Gaberdiel:2017oqg,Giribet:2018ada,Gaberdiel:2018rqv}.

So far we have considered the case of a single modulus turned on. In general we can turn on any combination of a constant $C_0$ and a constant and self-dual $C_2$. By a rescaling and rotation of the torus directions we can always align $C_2$ so that the non-vanishing components point in directions $67$ and $89$. Hence we are lead to consider a solution with
\begin{equation}
  C_0 = c_0 , \qquad
  C_2 = c_2 \bigl( e^6 \wedge e^7 + e^8 \wedge e^9 \bigr) , \qquad
  C_4 = -c_0 \, e^6 \wedge e^7 \wedge e^8 \wedge e^9 .
\end{equation}
This solution will have non-trivial RR field strengths
\begin{equation}
  F_3 = - C_0 H , \qquad
  F_5 = - C_2 \wedge H , \qquad
  F_7 = - C_4 \wedge H ,
\end{equation}
but does not carry any RR charges. In order to analyse this background we perform a TrT transformation in the directions $x^6$ and $x^7$ with a rotation angle of $\varphi$.\footnote{%
  It is worth noting that marginal deformations of the WZW model on $\Sphere^3$ have been studied in the past using transformations similar to TrT and TsT transformations~\cite{Hassan:1992gi,Giveon:1993ph,Forste:1994wp}. These differ from the RR deformations considered here because of their non-trivial effect on the $\Sphere^3$ metric.
} %
This results in a new background of the same type but where the RR potentials are now given by
\begin{equation}
  \begin{aligned}
    \tilde{C}_0 &= c_0 \cos\varphi + \sqrt{v} c_2 \sin\varphi , \\
    \tilde{C}_2 &= ( c_2 \cos\varphi - \sqrt{v} c_0 \sin\varphi ) ( \tilde{e}^6 \wedge \tilde{e}^7 + \tilde{e}^8 \wedge \tilde{e}^9 ) , \\
    \tilde{C}_4 &= - ( c_0 \cos\varphi + \sqrt{v} c_2 \sin\varphi )  \tilde{e}^6 \wedge \tilde{e}^7 \wedge \tilde{e}^8 \wedge \tilde{e}^9 ,
  \end{aligned}
\end{equation}
where $v$ is volume of the original $\Torus^4$. If we choose the angle $\varphi$ so that
\begin{equation}
  \tan \varphi = \frac{c_2}{\sqrt{v} c_0} ,
\end{equation}
the $\tilde{C}_2$ potential vanishes and we are left with a background of the type discussed earlier in this section, but with the value of the modulus and the $\AdS_3$ and $\Sphere^3$ radii taking the values
\begin{equation}
  \tilde{c}_0 = \frac{(c_2^2 + c_0^2)\sqrt{v}}{\sqrt{c_2^2 + v c_0^2}} , \qquad
  \tilde{R}^2 = \alpha' k \sqrt{ 1 + g_s^2 ( c_0^2 + c_2^2 ) } ,
\end{equation}
where $g_s$ is the string coupling before the TrT transformation.

\section{Fully backreacted geometries}
\label{sec:backreacted}

In the previous sections we have discussed how the closed string spectrum of the near-horizon geometry $\AdS_3 \times \Sphere^3 \times  \Torus^4$ is affected when the moduli of the background are turned on. We will now see how these moduli can be introduced in the full backreacted and asymptotically flat brane geometry. Following~\cite{Maldacena:1999mh} and~\cite{Dhar:1999ax} we will start with the D1/D5 system and introduce a B field by applying two TrT transformations, as well as adding a constant $B$ in order to cancel the resulting D3 charges. From the resulting background we can then obtain other F1/NS5 and D1/D5 backgrounds with non-trivial moduli by U duality. The corresponding near-horizon geometries are discussed in appendix~\ref{sec:general-ads3-s3-t4-bg}. Here we present the values of the parameters of the near-horizon solutions in terms of the physical brane charges.

\subsection{The D1/D5 system}
\label{sec:backreacted-d1-d5}

Let us start by writing down a type IIB supergravity solution corresponding to the standard D1/D5 system.
We consider a stack of $N_1$ D1 branes stretched along the directions\footnote{
  In order to not confuse the coordinates for the D1/D5 system with those of $\AdS_3$, we use the coordinates $\BBt$, $\BBx$ and $\BBr$ in the asymptotically flat geometry, and reserve $t$, $z$ and $r$ in the near-horizon limit.
}
$\BBt$ and $\BBx$, and a stack of $N_1$ D5 branes along $\BBt$, $\BBx$, $x^6$, $x^7$, $x^8$ and $x^9$, where the last four directions are compactified on a $\Torus^4$. The metric of the full D1/D5 system is given by
\begin{equation}\label{eq:D1-D5-metric}
  \begin{aligned}
    ds_{\text{D1/D5}}^2 &= (f_1f_5)^{-1/2} \bigl( - d\BBt^2 + d\BBx^2 \bigr) + ( f_1 f_5 )^{1/2} \bigl( d\BBr^2 + \BBr^2 ds_{\Sphere^3}^2 \bigr)
    \\ &\qquad\qquad
    + \Bigl( \frac{f_1}{f_5} \Bigr)^{1/2} \bigl( dx_6^2 + dx_7^2 + dx_8^2 + dx_9^2 \bigr) ,
  \end{aligned}
\end{equation}
where
\begin{equation}
  f_1 = 1 + \frac{\alpha'\nu_1}{\BBr^2} , \qquad
  f_5 = 1 + \frac{\alpha'\nu_5}{\BBr^2} .
\end{equation}
The D1/D5 system has a non-trivial dilaton
\begin{equation}
  \Phi = \frac{1}{2} \log \frac{f_1}{f_5} ,
\end{equation}
and is supported by a RR three-form 
\begin{equation}
  F_3 = - d f_1^{-1} \wedge d\BBt \wedge d\BBx + 2\alpha'\nu_5 \Omega_{\Sphere^3} .
\end{equation}
Note that in the asymptotic region, where $\BBr \to \infty$, the metric~\eqref{eq:D1-D5-metric} becomes flat, with the three sphere having unit radius, and the $\Torus^4$ having unit volume. Furthermore, the dilaton is normalised so that it vanishes asymptotically.

The corresponding D1 and D5 charges are give by
\begin{equation}
  \begin{aligned}
    Q_{\text{D5}} = \frac{1}{2\kappa_0^2} \int_{\Sphere^3} F_3 &= \frac{\nu_5}{(2\pi)^5(\alpha')^2} = \mu_5 N_5, \\
    Q_{\text{D1}} = \frac{1}{2\kappa_0^2} \int_{\Sphere^3 \times \Torus^4} *F_3 &= \frac{\nu_1}{2\pi\alpha'} = \mu_1 N_1 ,
  \end{aligned}
\end{equation}
where $\mu_p = (2\pi)^{-p} (\alpha')^{-(p+1)/2}$ is the charge density of the D$p$ brane. From this we find that the parameters $\nu_1$ and $\nu_5$ are related to the number of branes by
\begin{equation}
  \nu_1 = N_1 , \qquad
  \nu_5 = N_5 .
\end{equation}

\paragraph{Near-horizon geometry.}

In the near-horizon limit, the D1/D5 system geometry becomes $\AdS_3 \times \Sphere^3 \times \Torus^4$. The $\AdS_3$ and $\Sphere^3$ radii $R$ and the dilaton and $\Torus^4$ volume are now given by
\begin{equation}
  R^2 = \alpha' \sqrt{N_1 N_5} = \alpha' e^{\Phi} N_5 , \qquad
  e^{2\Phi} = \Vol(\Torus^4) = \frac{N_1}{N_5} ,
\end{equation}
while the other parameters are turned off.

\subsection{Turning on a B field in the D1/D5 system}
\label{sec:backreacted-d1-d5-b}

To turn on a B field on $\Torus^4$ we employ the same strategy as was discussed previously in the near-horizon geometry: we perform a TrT${}^2$ transformation, and add a constant B which can be adjusted so that there is no D3 charge.

The resulting metric takes to form
\begin{multline}
  d\tilde{s}^2_{\text{D1/D5}} = (f_1f_5)^{-1/2} \bigl( - d\BBt^2 + d\BBx^2 \bigr) + (f_1 f_5)^{1/2} \bigl( d\BBr^2 + \BBr^2 ds_{\Sphere^3}^2 \bigr) \\
  + (f_1 f_5)^{1/2} f_{\varphi}^{-1} \bigl( d\tilde{x}_6^2 + d\tilde{x}_7^2 \bigr) + (f_1 f_5)^{1/2} f_{\psi}^{-1} \bigl( d\tilde{x}_8^2 + d\tilde{x}_9^2 )\bigr) ,
\end{multline}
and the dilaton is given by
\begin{equation}
  \tilde{\Phi} = \frac{1}{2} \log\frac{f_1 f_5}{f_{\varphi} f_{\psi}} ,
\end{equation}
where
\begin{equation}
  \begin{aligned}
    f_{\varphi} &= 1 + \frac{\alpha' \nu_{\varphi}}{\BBr^2} , \quad &
    \nu_{\varphi} &= \nu_5 \cos^2\varphi + \nu_1 \sin^2\varphi ,
    \\
    f_{\psi} &= 1 + \frac{\alpha' \nu_{\psi}}{\BBr^2} , &
    \nu_{\psi} &= \nu_5 \cos^2\psi + \nu_1 \sin^2\psi .
  \end{aligned}
\end{equation}
Introducing the three forms
\begin{equation}
  \begin{aligned}
    K_3 &= -d f_5^{-1} \wedge d\BBt \wedge d\BBx + 2\alpha' \nu_1 \Omega_{\Sphere^3} , \\
    \tilde{K}_3 &= -d f_1^{-1} \wedge d\BBt \wedge d\BBx + 2\alpha' \nu_5 \Omega_{\Sphere^3} , \\
  \end{aligned}
\end{equation}
we can write the other non-trivial background fields as
\begin{equation}
  \begin{gathered}
    \tilde{F}_3 = \cos\varphi \cos\psi \tilde{K}_3 - \sin\varphi \sin\psi K_3 , 
    \\
    \begin{aligned}
      \tilde{F}_5 = &
      - f_{\varphi}^{-1} \bigl( f_5 \cos\varphi \sin\psi K_3 + f_1 \sin\varphi \cos\psi \tilde{K}_3 \bigr) \wedge d\tilde{x}^6 \wedge d\tilde{x}^7
      \\ &\qquad
      - f_{\psi}^{-1} \bigl( f_5 \sin\varphi \cos\psi K_3 + f_1 \cos\varphi \sin\psi \tilde{K}_3 \bigr) \wedge d\tilde{x}^8 \wedge d\tilde{x}^9 .
    \end{aligned}
    \\
    \begin{aligned}
      \tilde{B} =
      &-\bigl( f_{\varphi}^{-1} (f_1 - f_5) \cos\varphi \sin\varphi - b_{67} \bigr) d\tilde{x}^6 \wedge d\tilde{x}^7
      \\ &\qquad
      -\bigl( f_{\psi}^{-1} (f_1 - f_5) \cos\psi \sin\psi - b_{89} \bigr) d\tilde{x}^8 \wedge d\tilde{x}^9
    \end{aligned}
  \end{gathered}
\end{equation}
Let us now calculate the various D$p$-brane charges carried by this solution, starting with the D3 charges, which are given by
\begin{equation}
  Q_{\text{D3}}^{67} = \frac{1}{2\kappa_0^2} \int_{\Sphere^3 \times \Torus^2_{89}} \bigl( \tilde{F}_5 - \tilde{B} \wedge \tilde{F}_3 \bigr) , \qquad
  Q_{\text{D3}}^{89} = \frac{1}{2\kappa_0^2} \int_{\Sphere^3 \times \Torus^2_{67}} \bigl( \tilde{F}_5 - \tilde{B} \wedge \tilde{F}_3 \bigr) .
\end{equation}
These charge densities are well-defined in the sense that they are given in terms of globally defined forms, and are independent of the transverse radial coordinate. Performing the integrals we get
\begin{equation}
  \begin{aligned}
    \mu_3^{-1} Q_{\text{D3}}^{67} &= \nu_1 \sin\varphi ( b_{89} \sin\psi - \cos\psi ) - \nu_5 \cos\varphi ( b_{89} \cos\psi + \sin\psi ) , \\
    \mu_3^{-1} Q_{\text{D3}}^{89} &= \nu_1 \sin\psi ( b_{67} \sin\varphi - \cos\varphi ) - \nu_5 \cos\psi ( b_{67} \cos\varphi + \sin\varphi ) .
  \end{aligned}
\end{equation}
The vanishing of the D3-brane charges then leads to
\begin{equation}\label{eq:b67-b89-no-D3}
  b_{67} = \frac{\nu_1 \cos\varphi \sin\psi + \nu_5 \sin\varphi \cos\psi}{\nu_1 \sin\varphi \sin\psi - \nu_5 \cos\varphi \cos\psi} , \qquad
  b_{89} = \frac{\nu_1 \sin\varphi \cos\psi + \nu_5 \cos\varphi \sin\psi}{\nu_1 \sin\varphi \sin\psi - \nu_5 \cos\varphi \cos\psi} .
\end{equation}
From now on we will impose the above relations.

The D1 and D5 charges of the transformed background are given by
\begin{equation}\label{eq:D5-charge-D1-D5-B}
  Q_{\text{D5}} = \frac{1}{2\kappa_0^2} \int_{\Sphere^3} \tilde{F}_3 = \mu_5 \tilde{\nu} = \mu_5 \tilde{N}_5,
\end{equation}
and
\begin{equation}\label{eq:D1-charge-D1-D5-B}
  Q_{\text{D1}}
  = \frac{1}{2\kappa_0^2} \int_{\Sphere^3 \times \Torus^4} \bigl( *\tilde{F}_3 + \tilde{B} \wedge \tilde{F}_5 - \tfrac{1}{2} \tilde{B} \wedge \tilde{B} \wedge \tilde{F}_3 \bigr)
  = \frac{\mu_1 \nu_1 \nu_5}{\tilde{\nu}} = \mu_1 \tilde{N}_1 ,
\end{equation}
where
\begin{equation}\label{eq:tilde-nu-def}
  \tilde{\nu} = \nu_5 \cos\varphi \cos\psi - \nu_1 \sin\varphi \sin\psi.
\end{equation}
The above relations can be used to express the parameters $\nu_1$ and $\nu_5$ in terms of the physical quantities $\tilde{N}_1$ and $\tilde{N}_5$.\footnote{Note that the TrT transformation changes the charge quantisation condition so that we now should express the parameter $\nu_1$ and $\nu_5$ in terms of new integer charges $\tilde{N}_1$ and $\tilde{N}_5$.}

\paragraph{Near-horizon geometry.}

In the near-horizon limit, the transformed background is still given by $\AdS_3 \times \Sphere^3 \times \Torus^4$, with the $\AdS_3$ and $\Sphere^3$ radii, dilaton and $\Torus^4$ volume given by
\begin{equation}
  R^2 = \alpha'\sqrt{\nu_1 \nu_5} = \alpha' \sqrt{\tilde{N}_1 \tilde{N}_5} , \qquad
  e^{2\tilde{\Phi}} = \Vol(\Torus^4) = \frac{\nu_1 \nu_5}{\nu_{\varphi} \nu_{\psi}} = \frac{\tilde{N}_1}{\tilde{N}_5 + \tilde{N}_1 \sin^2( \varphi + \psi)} .
\end{equation}
The near-horizon B field can be written as $B = b(e^6 \wedge e^7 + e^8 \wedge e^9)$, with
\begin{equation}
  b = - \sqrt{\frac{\tilde{N}_1}{\tilde{N}_5}} \sin(\varphi + \psi) .
\end{equation}
We can the write the radius $R$ and the $\Torus^4$ volume as
\begin{equation}
  R^2 = \alpha' e^{\tilde{\Phi}} \tilde{N}_5 \sqrt{ 1 + b^2 } , \qquad
  \Vol(\Torus^4) = \frac{1}{1 + b^2} \frac{\tilde{N}_1}{\tilde{N}_5} .
\end{equation}
In this form the dependence on the moduli becomes manifest.

\subsection{The F1/NS5 system with a RR two form}
\label{sec:backreacted-f1-ns5-c2}

Let us now apply S duality to the TrT transformed D1/D5 system. We find the metric
\begin{equation}
  \begin{aligned}
    ds^2 &= \frac{\sqrt{f_{\varphi} f_{\psi}}}{f_1 f_5} \bigl( -d\BBt^2 + d\BBx^2 \bigr) + \sqrt{f_{\varphi} f_{\psi}} \bigl( d\BBr^2 + \BBr^2 ds_{\Sphere^3}^2 \bigr) \\
    & \qquad + \sqrt{\frac{f_{\psi}}{f_{\varphi}}} \bigl( dx_6^2 + dx_7^2 \bigr) + \sqrt{\frac{f_{\varphi}}{f_{\psi}}} \bigl( dx_8^2 + dx_9^2 \bigr) .
  \end{aligned}
\end{equation}
The dilaton is given by
\begin{equation}
  \Phi = \frac{1}{2} \log \frac{f_{\varphi} f_{\psi}}{f_1 f_5} .
\end{equation}
The geometry is supported by the NSNS three form
\begin{equation}
  H = \cos\varphi \cos\psi \tilde{K}_3 - \sin\varphi \sin\psi K_3 .
\end{equation}
There is also a RR five form
\begin{equation}
  \begin{split}
    F_5 = &
    - f_{\varphi}^{-1} \bigl( f_5 \cos\varphi \sin\psi K_3 + f_1 \sin\varphi \cos\psi \tilde{K}_3 \bigr) \wedge d\tilde{x}^6 \wedge d\tilde{x}^7
    \\ &\qquad
    - f_{\psi}^{-1} \bigl( f_5 \sin\varphi \cos\psi K_3 + f_1 \cos\varphi \sin\psi \tilde{K}_3 \bigr) \wedge d\tilde{x}^8 \wedge d\tilde{x}^9 .
  \end{split}
\end{equation}
as well as the RR two-form potential
\begin{equation}
  \begin{split}
    C_2 =
    &\bigl( f_{\varphi}^{-1} (f_1 - f_5) \cos\varphi \sin\varphi - b_{67} \bigr) d\tilde{x}^6 \wedge d\tilde{x}^7
    \\ &\qquad
    \bigl( f_{\psi}^{-1} (f_1 - f_5) \cos\psi \sin\psi - b_{89} \bigr) d\tilde{x}^8 \wedge d\tilde{x}^9 .
  \end{split}
\end{equation}
This potential does not lead to any D1-brane or D5-brane charges. To see that there also is no D3-brane charges we compute
\begin{equation}
  \frac{1}{2\kappa_0^2} \int_{\Sphere^3 \times \Torus^2} \bigl( F_5 + C_2 \wedge H \bigr) = 0 .
\end{equation}
Hence, the only non-vanishing charges are the NS5 charge
\begin{equation}\label{eq:F1-NS5-charge-QNS5}
  Q_{\text{NS5}} = \frac{1}{2\kappa_0^2} \int_{\Sphere^3} H = \mu_5 \tilde{\nu} .
\end{equation}
and the F1 charge
\begin{equation}\label{eq:F1-NS5-charge-QF1}
  Q_{\text{F1}} = \frac{1}{2\kappa_0^2} \int_{\Sphere^3 \times \Torus^4} \bigl( e^{-2\Phi} * H + C_0 F_7 + \tfrac{1}{2} C_2 \wedge C_2 \wedge H \bigr)
  = \frac{\mu_1 \nu_1 \nu_5}{\tilde{\nu}} ,
\end{equation}
Note that these charges take the same values as the D5 and D1 charges in equations~\eqref{eq:D5-charge-D1-D5-B} and~\eqref{eq:D1-charge-D1-D5-B}.

\paragraph{Near-horizon geometry.}
In the near-horizon limit, the $\AdS_3$ and $\Sphere^3$ radii are given by
\begin{equation}
  R^2 = \alpha' \sqrt{\nu_1 \nu_5} = \alpha' \tilde{N}_5 \sqrt{ 1 + \frac{\tilde{N}_1}{\tilde{N}_5} \sin^2 ( \varphi + \psi ) } ,
\end{equation}
where $\tilde{N}_1$ and $\tilde{N}_5$ now count the number of fundamental strings and NS5 branes.
The dilaton takes the form
\begin{equation}
  e^{2\Phi} = \frac{\nu_{\varphi} \nu_{\psi}}{\nu_1 \nu _5} = \frac{\tilde{N}_5}{\tilde{N}_1} \left( 1 + \frac{\tilde{N}_1}{\tilde{N}_5} \sin^2( \varphi + \psi ) \right) .
\end{equation}
The volume of the torus $\Vol(\Torus^4) = 1$ is constant in the full backreacted geometry and thus remains the same in the near horizon limit.

The RR two-form potential can be written as $C_2 = c_2 ( e^6 \wedge e^7 + e^8 \wedge e^9 )$ with
\begin{equation}
  c_2 = \frac{\tilde{N}_1}{\tilde{N}_5} \frac{\sin(\varphi+\psi)}{\sqrt{ 1 + \frac{\tilde{N}_1}{\tilde{N}_5} \sin^2 ( \varphi + \psi ) } } .
\end{equation}
We can the write the radius $R$ and the torus volume directly in terms of physical parameters as
\begin{equation}
  R^2 = \alpha' \tilde{N}_5 \sqrt{ 1 + e^{2\Phi} c_2^2 } , \qquad
  \Vol(\Torus^4) = \frac{e^{2\Phi}}{ 1 + e^{2\Phi} c_2^2 } \frac{\tilde{N}_1}{\tilde{N}_5} .
\end{equation}

\subsection{The F1/NS5 system with a RR scalar and four form}
\label{sec:backreacted-f1-ns5-c0}

We finally apply T duality transformations along directions $x^6$ and $x^7$. We then obtain the metric
\begin{equation}
  \begin{aligned}
    ds^2 &= \frac{\sqrt{f_{\varphi} f_{\psi}}}{f_1 f_5} \bigl( -d\BBt^2 + d\BBx^2 \bigr) + \sqrt{f_{\varphi} f_{\psi}} \bigl( d\BBr^2 + \BBr^2 ds_{\Sphere^3}^2 \bigr) \\
    & \qquad\qquad + \sqrt{\frac{f_{\psi}}{f_{\varphi}}} \bigl( dx_6^2 + dx_7^2 + dx_8^2 + dx_9^2 \bigr) .
  \end{aligned}
\end{equation}
The NSNS field strength remains the same as in the previous case,
\begin{equation}
  H = \cos\varphi \cos\psi \tilde{K}_3 - \sin\varphi \sin\psi K_3 .
\end{equation}
The turned on modulus is encoded in the RR scalar, which takes the form
\begin{equation}
  C_0 = c_0
  - \frac{1}{2} \frac{\nu_5 - \nu_1}{\nu_{\varphi}} \frac{\sin 2\varphi}{f_{\varphi}} ,
\end{equation}
where
\begin{equation}\label{eq:F1-NS5-RR-scalar}
  c_0 = -\frac{\nu_1 \nu_5}{\nu_{\varphi} \tilde{\nu}} \sin(\varphi+\psi) .
\end{equation}
Also the RR four-form potential is turned on
\begin{equation}
  \begin{aligned}
    C_4 &= \Bigl( -c_0 + \frac{\nu_5 - \nu_1}{\nu_{\varphi}} \frac{\nu_5 \cos^3\varphi \sin\psi - \nu_1 \sin^3\varphi \cos\psi}{f_{\psi}}
    \Bigr) e^6 \wedge e^7 \wedge e^8 \wedge e^9 .
  \end{aligned}
\end{equation}
Finally, the RR three-form field strength is given by
\begin{equation}
  F_3 = \frac{f_5}{f_{\varphi}} \cos\varphi \cos\psi K_3 + \frac{f_1}{f_{\psi}} \sin\varphi \cos\psi \tilde{K}_3 .
\end{equation}
While this solution has several non-vanishing RR field strengths, all the D brane charges are zero. The F1 and NS5 charges are the same as in equations~\eqref{eq:F1-NS5-charge-QF1} and ~\eqref{eq:F1-NS5-charge-QNS5}.

\paragraph{Near-horizon geometry.}
In the near-horizon limit, the $\AdS_3$ and $\Sphere^3$ radii are given by
\begin{equation}
  R^2 = \alpha' \sqrt{\nu_1 \nu_5} = \alpha' \tilde{N}_5 \sqrt{ 1 + \frac{\tilde{N}_1}{\tilde{N}_5} \sin^2 ( \varphi + \psi ) } ,
\end{equation}
and the dilaton and torus volume by
\begin{equation}
  e^{\Phi} = \frac{\nu_{\varphi}^2}{\nu_1 \nu_5} , \qquad
  \Vol(\Torus^4) = \frac{\nu_{\varphi}}{\nu_{\psi}} .
\end{equation}
The RR scalar and four form potential are constant in the near-horizon limit,
\begin{equation}
  C_0 = c_0 , \qquad
  C_4 = -c_0 e^6 \wedge e^7 \wedge e^8 \wedge e^9 ,
\end{equation}
where $c_0$ is defined in equation~\eqref{eq:F1-NS5-RR-scalar}. In terms of this parameter we can write
\begin{equation}
  R^2 = \alpha' \tilde{N}_5 \sqrt{ 1 + e^{2\Phi} c_0^2 } , \qquad
    \Vol(\Torus^4) = \frac{e^{2\Phi}}{ 1 + e^{2\Phi} c_0^2 } \frac{\tilde{N}_1}{\tilde{N}_5} .
\end{equation}

\subsection{The D1/D5 system with RR scalar and four form}
\label{sec:backreacted-d1-d5-c0}

We can finally make an S duality transformation to get back to a D1/D5 background with a RR scalar and four-form potential. The resulting metric takes the form
\begin{equation}
  \begin{aligned}
    ds^2 &= \frac{1}{\tilde{\nu}} \frac{\sqrt{\tilde{f}f_{\psi}}}{f_1 f_5} ( -d\BBt^2 + d\BBx^2 )
    + \frac{\sqrt{\tilde{f} f_{\psi}}}{\tilde{\nu}} ( d\BBr^2 + \BBr^2 ds_{\Sphere^3}^2 )
    \\ &\qquad\qquad
    + \frac{1}{\tilde{\nu}}\sqrt{\frac{\tilde{f}}{f_{\psi}}} ( dx_6^2 + dx_7^2 + dx_8^2 + dx_9^2 ) ,
  \end{aligned}
\end{equation}
where $\tilde{\nu}$ was introduced in~\eqref{eq:tilde-nu-def} and
\begin{equation}
  \tilde{f} = \nu_5^2 \cos^2\psi f_1 + \nu_1^2 \sin^2\psi f_5 .
\end{equation}
The dilaton and the $\Torus^4$ volume are now given by
\begin{equation}
  e^{\Phi} = \frac{\tilde{f}}{\sqrt{f_1 f_5} \tilde{\nu}^2 } , \qquad
  \Vol(\Torus^4) = \frac{\tilde{f}}{f_{\psi} \tilde{\nu}^2} .
\end{equation}
The background is supported by the RR three-form
\begin{equation}
  F_3 = \frac{\tilde{\nu}}{\nu_{\psi}} \frac{f_{\psi}}{f_1 f_5 \tilde{f}} \, d\tilde{f} \wedge d\BBt \wedge \BBx + 2\alpha' \tilde{\nu} \Omega_{\Sphere^3} .
\end{equation}
There is furthermore a B field
\begin{equation}
  B = \frac{(f_1^{-1} - f_5^{-1}) r^2 \sin(2\psi)}{2\alpha' \tilde{\nu}} d\BBt \wedge d\BBx ,
\end{equation}
a RR scalar
\begin{equation}
  C_0 = c_0 + \frac{r^2\tilde{\nu}^2 (f_1 - f_5) \sin(2\psi)}{2\alpha' \nu_{\psi} \tilde{f}} ,
\end{equation}
where
\begin{equation}\label{eq:D1-D5-C0-c0-def}
  c_0 = \frac{\tilde{\nu}}{\nu_{\psi}} \sin(\varphi + \psi) ,
\end{equation}
and a RR four form
\begin{equation}
  C_4 = B \wedge C_2 + \Bigl( c_0 + \frac{r^2 \tilde{\nu} (f_1 - f_5) ( \nu_5 \sin\varphi \cos^3\psi - \nu_1 \cos\varphi \sin^3\psi )}{ \alpha' \nu_{\psi} \tilde{f}} \Bigr) e^6 \wedge e^7 \wedge e^8 \wedge e^9.
\end{equation}
The non-vanishing charges of the background are the D5 and D1 charges, which take the same values as in equations~\eqref{eq:D5-charge-D1-D5-B} and~\eqref{eq:D1-charge-D1-D5-B}

\paragraph{Near-horizon geometry.}

In the near-horizon limit the $\AdS_3$ and $\Sphere^3$ radii, the dilaton and the $\Torus^4$ volume are given by
\begin{equation}
  R^2 = \frac{\alpha' \nu_{\psi} \sqrt{\nu_1 \nu_5}}{\tilde{\nu}} = \alpha' e^{\Phi} \tilde{N}_5, \qquad
  e^{\Phi} = \frac{\sqrt{\nu_1 \nu_5} \nu_{\psi}}{\tilde{\nu}^2 } , \qquad
  \Vol(\Torus^4) = \frac{\nu_1 \nu_5}{\tilde{\nu}^2 } = \frac{\tilde{N}_1}{\tilde{N}_5} .
\end{equation}
Note that both the radius and the torus volume are independent of the modulus. The RR scalar and four form are given by
\begin{equation}
  C_0 = c_0 , \qquad
  C_4 = c_0 \, e^6 \wedge e^7 \wedge e^8 \wedge e^9 ,
\end{equation}
where $c_0$ is defined in~\eqref{eq:D1-D5-C0-c0-def}.

\section{Conclusions}
\label{sec:conclusions}

In this paper we have determined how the energies of closed perturbative strings on 
$\AdS_3\times \Sphere^3\times \Torus^4$ with zero winding and momentum on the torus depend on the 20 moduli of the string theory. We focused on backgrounds which are near-horizon limits of D1/D5- and F1/NS5-branes  and found that in both cases only four of the 20 moduli have a non-trivial effect on the Green-Schwarz action, and as a result have an effect on the energies of $\mathcal{H}_{(0,0)}$ strings.

In the near-horizon limit of  D1/D5-branes one of the consequential moduli is the closed string coupling constant whose only effect (at small $g_s$) is to change the radius of curvature $R$. The remaining three consequential moduli come from the self-dual part of the NSNS two-form. When $B^+\neq 0$, the GS action involves couplings to the RR five-form field strength, in addition to the three form which supports the geometry. We showed that this new action is TrT-dual to the original action with $B^+ = 0$, and that closed strings in the $B^+ \neq 0$ background have the same energies as strings in the $B^+ = 0$ background with twisted boundary conditions~\eqref{eq:tw-per-cond}. The twisting is proportional to the amount of winding and momentum on $\Torus^4$ (as well as to  $B^+$), and so in the case of $\mathcal{H}_{(0,0)}$ strings the only change to the energies comes from a change in the radius of curvature~\eqref{eq:rad-d1d5-bmod}.

In the near-horizon limit of F1/NS5-branes, the four consequential moduli are the RR scalar $C_0$ and self-dual part of the two-form $C_2^+$. The GS action in a background with a general combination of these moduli turned on is TrT-dual to the GS action with just $C_0$ turned on. As a result, for the $\mathcal{H}_{(0,0)}$ spectrum we may investigate the effect of just the $C_0$ modulus: turning on the other RR moduli is encoded into a change in the radius of curvature. The GS action depends on gauge-invariant RR field strengths
\begin{equation}
F_{p+1}=dC_p-C_{p-2}\wedge H .
\end{equation}
As a result, a constant $C_0$ in the $\AdS_3$ geometry supported by NSNS flux induces a RR three-form flux, and the GS action is the same as the so-called mixed flux $\AdS_3$ backgrounds~\cite{Cagnazzo:2012se,Hoare:2013ida,Lloyd:2014bsa}. The interpretation is however very different. The mixed-flux backgrounds correspond to the near-horizon limit of $(p,q)$-strings and 5-branes, while the background we investigated in this paper is a marginal deformation of the pure NSNS-flux background, and carries only F1 and NS5 charges. When $C_0$ vanishes, the action is the GS analogue of the Neveu-Schwarz-Ramond theory considered by Maldacena and Ooguri~\cite{Maldacena:2000hw}. With general RR moduli turned on the  radius of curvature is
\begin{equation}
R^2=\alpha' k \sqrt{1+g_s^2c^2}\,\qquad\mbox{where}\qquad c^2=C_0^2+\tfrac{1}{2}(C^+_2)^2 .
\label{eq:mixedf-radius}
\end{equation}

Integrable methods have provided exact-in-$R$ results for determining the energies of  $\mathcal{H}_{(0,0)}$ states at small $g_s$~\cite{Babichenko:2009dk,Sax:2012jv,Borsato:2013qpa}. In particular, in the near horizon limit of the D1/D5-brane theory, the exact-in-$R$ 2-to-2 worldsheet S matrix is fixed by symmetries alone~\cite{Borsato:2014exa,Borsato:2014hja} at the point in moduli space given in equation~\eqref{eq:orig-mod-d1-d5}.\footnote{%
  More accurately, a number of so-called dressing factors are not fixed by symmetries alone, but can be found using unitarity and crossing symmetry~\cite{Janik:2006dc} of the theory~\cite{Borsato:2013hoa,Borsato:2016kbm,Borsato:2016xns}. In the mixed flux case, the dressing factors are currently only known at the one-loop level~\cite{Babichenko:2014yaa}.%
}
\emph{A posteriori} the S matrix turns out to satisfy the Yang-Baxter equation, and so the complete worldsheet scattering problem reduces to combinations of pairwise scattering described by the S matrix~\cite{Borsato:2014exa,Borsato:2014hja}. The exact-in-$R$ $\mathcal{H}_{(0,0)}$ spectrum can then be obtained using Bethe Ansatz methods at small $g_s$. It consists of the BMN vacuum~\cite{Berenstein:2002jq}, on top of which magnon-like  operators act to create excited states. The multiplicities of these creation operators is determined by the BMN spectrum~\cite{Berenstein:2002jq}. Each magnon has mass $m=0,1$ or $-1$ and carries a momentum $p$, whose value is determined as a solution of the Bethe Equations~\cite{Borsato:2016kbm,Borsato:2016xns}. The energy of magnon is fixed through a $p$-dependent shortening condition~\cite{David:2008yk,Borsato:2014exa}
\begin{equation}
  \label{eq:rr-disp-rel}
  E(p)=\sqrt{m^2+4h^2(R)\sin^2\left(\tfrac{p}{2}\right)},
\end{equation}
and the energy of the state is given by the sum of the energies of the individual magnons.\footnote{Incorporating wrapping effects~\cite{Ambjorn:2005wa} into this construction remains to be fully understood~\cite{Abbott:2015pps}, but it appears likely that this will be possible~\cite{BSTtoappear}.}  $R$ enters the above dispersion relation through the function $h(R)$, whose explicit form is not fixed by integrability, much like is the case in $\AdS_4\times \CP^3$~\cite{Nishioka:2008gz,Gaiotto:2008cg}. As in higher-dimensional examples of integrable holography, $h$ determines the strength of the integrable interactions and at large $R$
\begin{equation}
h(R)= \frac{R^2}{2\pi}+\dots.
\end{equation}
Determining $h$, perhaps along the lines used in $\AdS_4\times \CP^3$~\cite{Gromov:2014eha}, remains an important problem.\footnote{%
  Perturbative world-sheet corrections to the dispersion relation in $\AdS_3 \times \Sphere^3 \times \Torus^4$ and $\AdS_3 \times \Sphere^3 \times \Sphere^3 \times \Sphere^1$ were calculated in~\cite{Sundin:2012gc,Sundin:2014ema}.%
}
The analysis carried out in section~\ref{sec:d1-d5-moduli}, shows that the world-sheet theory continues to be integrable across the 20 dimensional moduli space. The four consequential moduli modify the integrable structure of the $\mathcal{H}_{(0,0)}$ spectrum in a minimal way: they just changing the value of $R$~\eqref{eq:rad-d1d5-bmod}, which changes the value of $h(R)$. Since the BEs are valid for all values of $h$, we retain complete control over the spectrum across the whole moduli space as longs as $g_s$ is small. For example, in~\cite{Baggio:2017kza}, the half-BPS spectrum of the theory was found and shown to match the supergravity results of~\cite{deBoer:1998kjm}. The derivation in~\cite{Baggio:2017kza} is exact in $R$, and combined with the present results, proves that the half-BPS spectrum does not change as we move around moduli space, in agreement with the non-renormalization theorem of~\cite{Baggio:2012rr}.

Integrable methods have also been  used to find the $\mathcal{H}_{(0,0)}$ spectrum in mixed-flux backgrounds~\cite{Cagnazzo:2012se,Hoare:2013ida,Babichenko:2014yaa,Lloyd:2014bsa}.\footnote{Semi-classical strings in such backgrounds have also been studied in~\cite{Hernandez:2014eta,David:2014qta,Banerjee:2015qeq,Banerjee:2015bia,Hernandez:2018gcd}.}
However, to date it was not possible to study the pure F1/NS5 near-horizon geometry, because the off-shell supersymmetry algebra is \emph{not} centrally extended in that case~\cite{Lloyd:2014bsa}, and these central extensions are crucial in fixing the S matrix from symmetries~\cite{Beisert:2005tm,Arutyunov:2006ak}. However, this obstacle occurs only at the point in moduli space given in equation~\eqref{eq:orig-mod-f1-ns5}. As we showed in section~\ref{sec:f1-ns5-moduli}, when the RR moduli are non-zero the GS action becomes equivalent to the GS mixed-flux action, for which the central extensions are again non-zero.\footnote{It is worth noting that turning on such RR moduli is expected to desingularise the dual $\CFT_2$~\cite{Seiberg:1999xz}.} We can therefore use the exact-in-$R$ results of the mixed-flux background directly, upon re-expressing the parameters as in equation~\eqref{eq:mixedf-reexp-param}. The spectrum again consists of BMN magnon creation operators, but now with the dispersion relation~\cite{Hoare:2013ida,Hoare:2013lja,Lloyd:2014bsa}
\begin{equation}
E(p)=\sqrt{\left(m+\tfrac{k p}{2\pi}\right)^2+4h^2(R)\sin^2\left(\tfrac{p}{2}\right)},
\label{eq:mixed-flux-disp}
\end{equation}
where $k$ is the WZW level, the $\AdS_3$ radius is given in equation~\eqref{eq:mixedf-radius} and at large $R$, we have
\begin{equation}
h(R)= -\frac{g_s c k}{2\pi}+\dots.
\end{equation}
We arrive at a nice picture: the near-horizon F1/NS5-brane worldsheet theory is integrable with the strength of the integrable interactions governed by $c k$, and the free point corresponding to the GS version of the WZW model solved by~\cite{Maldacena:2000hw}.
By analogy with higher-dimensional holography, we may think of $c k$ as $\sqrt{\lambda}$, the analogue of the 't Hooft coupling constant, which should lead to a novel planar limit for the  near-horizon F1/NS5-brane theory. We intend to return to a more detailed investigation of this in the near future.

It is worth emphasizing that both in the D1/D5 and F1/NS5 backgrounds, our arguments show 
that the S matrix and hence the BEs are the same as the ones obtained in~\cite{Borsato:2014exa,Borsato:2014hja} and~\cite{Lloyd:2014bsa} respectively. 
The dependence on the moduli is contained entirely within the function $h$ that enters the dispersion relation
and the definition of the Zhukovsky variables.

Given the exact-in-$R$ nature of the integrable holographic methods now available across the whole moduli space, we believe the spectrum of the $\mathcal{H}_{(0,0)}$ sector is an ideal tool for investigating more precisely the relationship between strings on $\AdS_3$ and its $\CFT_2$ dual. It is for example striking that the dimensions of $\SymN$ states with zero winding and momentum also depend only on the four moduli of the $Z_2$-twisted sector, just as our $\mathcal{H}_{(0,0)}$  states do. Nonetheless, as the recent findings for the WZW theory at $k=1$ suggest~\cite{Gaberdiel:2017oqg,Giribet:2018ada,Gaberdiel:2018rqv}, the exact relationship to the $\SymN$ orbifold most likely needs to be revisited, and the $\mathcal{H}_{(0,0)}$ sector contains a wealth of non-protected information with which to test any such conjectures. It seems plausible that the more complete holographic dual will have to incorporate aspects of an effective Higgs branch $\CFT$~\cite{Witten:1997yu}, where integrability has been found~\cite{Sax:2014mea}. 

We also hope to extend the analysis carried out here to strings on $\AdS_3\times\Sphere^3\times\Sphere^3\times\Sphere^1$ backgrounds, since these too are known to be governed by integrable world-sheet theories~\cite{Babichenko:2009dk,OhlssonSax:2011ms,Borsato:2012ud,Borsato:2012ss,Sundin:2012gc,Abbott:2013mpa,Borsato:2015mma}. The moduli space is much smaller here~\cite{Gukov:2004ym}, and some conjectures for the $\CFT_2$ dual also exist~\cite{Gukov:2004ym,Tong:2014yna,Eberhardt:2017pty}.

\section*{Acknowledgements}

We would like to thanks Costas Bachas, Marcus Berg, Chris Hull, Oleg Lunin, A.W. Peet, Boris Pioline, Sanjaye Ramgoolam, Leonardo Rastelli, Rodolfo Russo and Konstantin Zarembo for interesting discussions, and Riccardo Borsato, Alessandro Sfondrini and Alessandro Torrielli for the many conversations on integrability and AdS3/CFT2. We would like to thank GGI for hosting the workshop "New Developments in AdS3/CFT2 Holography" where this work begun. B.S.\@ thanks the CERN Theory division and Helsinki Institute of Physics for hospitality during parts of this project. B.S.\@ acknowledges funding support from an STFC Consolidated Grant "Theoretical Physics at City University" ST/P000797/1. This work was supported by the ERC advanced grant No 341222.

\appendix

\section{Supergravity}

\subsection{Conventions}

We define Hodge duality by
\begin{equation}
  * ( e^{a_1} \wedge \dotsb \wedge e^{a_k} ) = \frac{1}{(d-k)!} \epsilon^{a_1\dotsc a_k}{}_{b_{k+1} \dotsc b_{d}} e^{b_{k+1}} \wedge \dotsb \wedge e^{b_d} ,
\end{equation}
where
\begin{equation}
  \epsilon^{0123456789} = +1 .
\end{equation}

For $\AdS_3$ we use the coordinate $t$, $z_1$ and $z_2$ with metric
\begin{equation}
  ds_{\AdS_3}^2 = - \biggl( \frac{1 + \frac{|z|^2}{4}}{1 - \frac{|z|^2}{4}} \biggr)^2 dt^2 + \biggl( \frac{1}{1 - \frac{|z|^2}{4}} \biggr)^2 |dz|^2 ,
\end{equation}
and for $\Sphere^3$ we use $y_3$, $y_4$ and $\phi$, with the metric
\begin{equation}
  ds_{\Sphere^3}^2 = \biggl( \frac{1 - \frac{|y|^2}{4}}{1 + \frac{|y|^2}{4}} \biggr)^2 d\phi^2 + \biggl( \frac{1}{1 + \frac{|y|^2}{4}} \biggr)^2 |dy|^2 .
\end{equation}
The corresponding unit volume forms are given by
\begin{equation}
  \Omega_{\AdS_3} = \frac{1 + \frac{|z|^2}{4}}{\bigl( 1 - \frac{|z|^2}{4} \bigr)^3} \, dt \wedge dz_1 \wedge dz_2 ,
  \qquad
  \Omega_{\Sphere_3} = \frac{1 - \frac{|y|^2}{4}}{\bigl( 1 + \frac{|y|^2}{4} \bigr)^3} \, dy_3 \wedge dy_4 \wedge d\phi .
\end{equation}
It is also useful to introduce the two forms
\begin{equation}
  \begin{aligned}
    \omega_{\AdS_3} &= \frac{1}{2} \frac{1}{\bigl(1 - \frac{|z|^2}{4}\bigr)^2} \, dt \wedge (z_2 dz_1 - z_1 dz_2 ) ,
    \\
    \omega_{\Sphere^3} &= \frac{1}{2} \frac{1}{\bigl(1 + \frac{|y|^2}{4}\bigr)^2} \, d\phi \wedge (y_4 dy_3 - y_3 dy_4 ) ,
  \end{aligned}
\end{equation}
which are define so that \emph{locally}
\begin{equation}
  \Omega_{\AdS_3} = d\omega_{\AdS_3} , \qquad
  \Omega_{\Sphere^3} = d\omega_{\Sphere^3} .
\end{equation}

\subsection{IIB supergravity}
\label{sec:iib-supergravity}

The action of IIB supergravity is given by\footnote{We follow the conventions of~\cite{Itsios:2012dc,Hamilton:2016ito}.}
\begin{equation}
  \begin{aligned}
    S_{\text{IIB}} &=
    \frac{1}{2\kappa_0^2}
    \int \!\! \sqrt{-g} \Bigl(
    e^{-2\Phi} \bigl( R + 4 (\partial \Phi)^2 - \tfrac{1}{12} H^2 \bigr) - \tfrac{1}{2} \bigl( F_1^2 + \tfrac{1}{3!} F_3^2 + \tfrac{1}{2\,5!} F_5^2 \bigr) \Bigr)
    \\ &\qquad
    - \frac{1}{4\kappa_0^2}
    \int C_4 \wedge H \wedge F_3 ,
  \end{aligned}
\end{equation}
where the gauge invariant field strengths are given in terms of gauge potentials by
\begin{equation}\label{eq:IIB-field-strengths}
  \begin{aligned}
    H &= dB , \quad &
    F_3 &= dC_2 - C_0 H , \quad &
    F_7 &= dC_6 - C_4 \wedge H , \\
    F_1 &= dC_0 , \quad &
    F_5 &= dC_4 - C_2 \wedge H , \quad &
    F_9 &= dC_8 - C_6 \wedge H ,
  \end{aligned}
\end{equation}
and satisfy
\begin{equation}
  * F_1 = F_9 , \quad
  * F_3 = - F_7 , \quad
  * F_5 = F_5 .
\end{equation}
The RR field strengths satisfy the Bianchi identities and equations of motions\footnote{%
  Note that the Bianchi identity for $F_n$ is the same as the equation of motion for $*F_n$.
}%
\begin{equation}
  \begin{gathered}
    dF_1 = 0, \quad
    dF_3 + F_1 \wedge H = 0, \quad
    dF_5 + F_3 \wedge H = 0 , \\
    dF_7 + F_5 \wedge H = 0 , \quad
    dF_9 + F_7 \wedge H = 0 .
  \end{gathered}
\label{eq:eoms-and-bianchis}
\end{equation}
The Bianchi identity and equation of motion for the NSNS three form $H$ are given by
\begin{equation}
  dH = 0 , \quad
  d\bigl( e^{-2\Phi} * H \bigr) - F_1 \wedge * F_3 - F_3 \wedge F_5 = 0 .
\end{equation}
The equations of motion for the metric $g_{\mu\nu}$ are given by
\begin{equation}
  R_{ab} - 2 \nabla_{a} \partial_{b} \Phi + \frac{1}{2} |H|^2_{ab} + \frac{1}{2} e^{2\Phi} \Bigl( | F_1|^2_{ab} + |F_3|^2_{ab} + \tfrac{1}{2} |F_5|^2_{ab} - \tfrac{1}{2} \eta_{ab} \bigl( |F_1|^2 + |F_3|^2 \bigr) \Bigr) = 0 ,
\end{equation}
where the symmetric contractions of the fields strengths are given, for example, by
\begin{equation}
  |H|^2_{ab} = \langle i_a H , i_b H \rangle = - * ( i_a H \wedge * i_b H ) .
\end{equation}
The equation of motion of the dilaton reads
\begin{equation}
  d*d\Phi - 2 |d\Phi|^2 + \tfrac{1}{2} |H|^2 - e^{2\Phi} \bigl( |F_1|^2 + \tfrac{1}{2} |F_3|^2 \bigr) = 0.
\end{equation}

From the equations of motion and Bianchi identities we define the conserved charges carried by D$p$ and NS5 branes as well as F1 strings. For example, the D3 charges is given by
\begin{equation}
  Q_{\text{D3}} = \frac{1}{2\kappa_0^2} \int_{\mathcal{M}_{\perp}} dF_5 + F_3 \wedge H ,
\end{equation}
where $\mathcal{M}_{\perp}$ is transverse to the branes. This charge is local and quantised, but is only invariant under small gauge transformations~\cite{Marolf:2000cb}. By partially integrating this in the radial direction away from the branes, this can be rewritten as an integral over a surface $\partial\mathcal{M}_{\perp}$ enclosing the branes. The exact form of the integrand will then depend on the exact form of the involved field strengths.

\subsection{T duality}
\label{sec:t-duality}

If we make a T-duality transformation along the direction $z$ then the NSNS fields transform as
\begin{equation}
  \begin{gathered}
    \tilde{B}_{\mu z} = - \frac{g_{z\mu}}{g_{zz}} , \qquad
    \tilde{B}_{\mu\nu} = B_{\mu\nu} + \frac{2}{g_{zz}} B_{z[\mu} g_{\nu]z} , \qquad
    \tilde{\Phi} = \Phi - \frac{1}{2} \log g_{zz} , \\
    \tilde{g}_{zz} = \frac{1}{g_{zz}} , \qquad
    \tilde{g}_{\mu z} = -\frac{B_{\mu z}}{g_{zz}} , \qquad
    \tilde{g}_{\mu\nu} = g_{\mu\nu} - \frac{1}{g_{zz}} \bigl( g_{\mu z} g_{\nu z} - B_{\mu z} B_{\nu z} \bigr) , 
  \end{gathered}
\end{equation}
and the RR fields transform as
\begin{equation}
  \begin{gathered}
    \tilde{C}^{(n)}_{\mu_1\dotsb\mu_{n-1} z} = C^{(n-1)}_{\mu_1 \dotsb \mu_{n-1}} - (-1)^n \frac{n-1}{g_{zz}} C^{(n-1)}_{z[\mu_1\dotsb\mu_{n-2}} g^{}_{\mu_{n-1}]z} , \\
    \tilde{C}^{(n)}_{\mu_1\dotsb\mu_n} = C^{(n+1)}_{\mu_1\dotsb\mu_n z} - n C^{(n-1)}_{[\mu_1\dotsb\mu_{n-1}} B^{}_{\mu_n]z}
    - (-1)^n \frac{n(n-1)}{g_{zz}} C^{(n-1)}_{z[\mu_1\dotsb\mu_{n-2}} B^{}_{\mu_{n-1}|z|} g^{}_{\mu_{n}]z} ,
  \end{gathered}
\end{equation}
where we have used a superscript to indicate the degree of the various forms.

\section{General \texorpdfstring{$\AdS_3 \times \Sphere^3 \times \Torus^4$}{AdS3 x S3 x T4} backgrounds}
\label{sec:general-ads3-s3-t4-bg}
In this appendix we will write down two general $\AdS_3 \times \Sphere^3 \times \Torus^4$ type IIB supergravity backgrounds, with the metric given by
\begin{equation}
  ds^2 = R^2 \bigl( ds_{\AdS_3}^2 + ds_{\Sphere^3} \bigr) + g_{ij} dx^i dx^j .
\end{equation}
We will focus on two special cases: backgrounds carrying only the RR charges of the D1/D5 system, and hence supported by a RR three form $F_3$ and its dual $F_7$, and backgrounds carrying F1 and NS5 charges, and thus sourcing the NSNS field strength $H$, and its dual.

There are a number of fields we can turn on that are compatible with the isometries of the above metric:
\begin{itemize}[noitemsep]
\item a constant B field on $\Torus^4$,
\item a constant two-form potential $C_2$ on $\Torus^4$,
\item the RR scalar $C_0$,
\item a constant dilaton $\Phi$,
\item a RR four-form potential $C_4$ on $\Torus^4$.
\end{itemize}
In the following two subsections we will write down general solutions with these fields turned on.

\subsection{RR backgrounds}
\label{sec:general-ads3-s3-t4-bg-rr}

Let us start by considering the case of backgrounds supported by RR flux. Imposing the equations of motion, as well as demanding the absence of D3 charges as well as any NSNS charges, we find that the dilaton is an arbitrary constant, that the constant B field is self dual
\begin{equation}
  B =
  b_{67} \bigl( e^6 \wedge e^7 + e^8 \wedge e^9 \bigr)
  + b_{68} \bigl( e^6 \wedge e^8 - e^7 \wedge e^9 \bigr)
  + b_{69} \bigl( e^6 \wedge e^9 + e^7 \wedge e^8 \bigr) ,
\end{equation}
and that the RR potentials are given by
\begin{equation}
  \begin{gathered}
    C_0 = c_0 ,
    \qquad C_8 = 0 ,
    \\
    C_2 = f \bigl( \omega_{\AdS_3} + \omega_{\Sphere^3} \bigr)
      + \tfrac{1}{2} c_{ij} \, e^i \wedge e^j
    \\
    C_4 = \tfrac{1}{2} f \bigl( \omega_{\AdS_3} + \omega_{\Sphere^3} \bigr) \wedge B + \tfrac{1}{2} C_2 \wedge B + c_0 \, e^6 \wedge e^7 \wedge e^8 \wedge e^9 ,
    \\
    C_6 = -f \bigl( \omega_{\AdS_3} + \omega_{\Sphere^3} \bigr) \wedge e^6 \wedge e^7 \wedge e^8 \wedge e^9 ,
  \end{gathered}
\end{equation}
with $c_{ji} = -c_{ij}$.
This corresponds to the field strengths
\begin{equation}
  \begin{gathered}
    F_1 = 0, \qquad
    F_9 = 0
    \\
    F_3 = f \bigl( \Omega_{\AdS_3} + \Omega_{\Sphere^3} \bigr) ,
    \qquad
    F_5 = f \bigl( \Omega_{\AdS_3} + \Omega_{\Sphere^3} \bigr) \wedge B ,
    \\
    F_7 = -f \bigl( \Omega_{\AdS_3} + \Omega_{\Sphere^3} \bigr) \wedge e^6 \wedge e^7 \wedge e^8 \wedge e^9 .
  \end{gathered}
\end{equation}
This results in a solution to the IIB supergravity equations of motion, provided the radius of $\AdS_3$ and $\Sphere^3$ is related to the other parameters by
\begin{equation}
  R^4 = \tfrac{1}{4} e^{2\Phi} f^2 \bigl( 1 + b_{67}^2 + b_{68}^2 + b_{69}^2 \bigr) .
\end{equation}
This solution includes a number of free parameters:
\begin{itemize}[noitemsep]
\item the metric of $\Torus^4$ -- 10 parameters,
\item the two-form potential $C_2$ on $\Torus^4$ -- 6 parameters,
\item the self-dual B field -- 3 parameters,
\item the dilaton $\Phi$ -- 1 parameter,
\item one linear combination of the RR potentials $C_0$ and $C_4$ -- 1 parameter,
\item the coefficient $f$ of the RR three-form field strength $F_3$ -- 1 parameter.
\end{itemize}
The last parameter, the coefficient $f$ in front of $F_3$, gives the radius of $\AdS_3$ and $\Sphere^3$. This leaves us with 21 parameters -- \emph{one more} than the expected number of moduli in the D1/D5 system. The reason for this is that the total volume of $\Torus^4$ is fixed by the attractor mechanism when the $\AdS_3 \times \Sphere^3 \times \Torus^4$ geometry is obtained in the near-horizon limit of the brane system. To find the volume of $\Torus^4$ in directly in the above supergravity solution, we need to impose flux quantisation of the field strengths $F_3$ and $F_7$. This gives us two additional constraints. One such constraint fixes $f$, and hence the radius $R$, in terms of the fluxes, and the second constraint determines the volume of $\Torus^4$.

Note that of the 21 remaining parameters, only the self-dual B field appears in the expressions for the field strengths. The Green-Schwarz string does not couple directly to the gauge potentials, but only to the gauge invariant field strengths.\footnote{From~\eqref{eq:IIB-field-strengths} we see that for $H=0$ the gauge invariant field strengths only depend on the derivatives of the gauge potentials.}
Hence, the only moduli the closed string spectrum in the D1/D5 system is sensitive to originates in the self-dual B field.

\subsection{NSNS backgrounds}
\label{sec:general-ads3-s3-t4-bg-nsns}

Let us now consider backgrounds that only carry F1 and NS5 charges. Again we impose the equations of motion and Bianchi identities, as well as the vanishing of all D-brane charges. We then find that the dilaton is constant and that the RR potentials are given by
\begin{equation}
  \begin{gathered}
    C_0 = c_0 , \qquad C_4 = -c_0 e^6 \wedge e^7 \wedge e^8 \wedge e^9 , \\
    C_2 = c_{67} ( e^6 \wedge e^7 + e^8 \wedge e^9 ) + c_{68} ( e^6 \wedge e^8 - e^7 \wedge e^9 ) + c_{69} ( e^6 \wedge e^9 + e^7 \wedge e^8 ) ,
  \end{gathered}
\end{equation}
while the B field is given by
\begin{equation}
  B = h \bigl( \omega_{\AdS_3} + \omega_{\Sphere^3} \bigr) + \tfrac{1}{2} b_{ij} \, e^i \wedge e^j ,
\end{equation}
with $b_{ji} = - b_{ij}$. The corresponding field strengths are given by
\begin{equation}
  H = h ( \Omega_{\AdS_3} + \Omega_{\Sphere^3} ) ,
\end{equation}
and
\begin{equation}
  \begin{gathered}
    F_1 = 0, \qquad F_9 = 0 , \\
    F_3 = -c_0 h ( \Omega_{\AdS_3} + \Omega_{\Sphere^3} ) , \\
    F_5 = -h ( \Omega_{\AdS_3} + \Omega_{\Sphere^3} ) \wedge C_2 , \\
    F_7 = + c_0 h ( \Omega_{\AdS_3} + \Omega_{\Sphere^3} ) \wedge e^6 \wedge e^7 \wedge e^8 \wedge e^9 .
  \end{gathered}
\end{equation}
The radius of $\AdS_3$ and $\Sphere^3$ is now given by
\begin{equation}
  R^2 = \tfrac{1}{4} h^2 \bigl( 1 + e^{2\Phi} ( c_0^2 + c_{67}^2 + c_{68}^2 + c_{69}^2 ) \bigr).
\end{equation}
Again, this solution has 22 free parameters:
\begin{itemize}[noitemsep]
\item the metric of $\Torus^4$ -- 10 parameters,
\item the self-dual two-form potential $C_2$ -- 3 parameters,
\item the B field -- 6 parameters,
\item the dilaton $\Phi$ -- 1 parameter,
\item one linear combination of the RR potentials $C_0$ and $C_4$ -- 1 parameter,
\item the coefficient $h$ of the NSNS three-form field strength $H$ -- 1 parameter.
\end{itemize}
As in the previous case, imposing charge quantisation lets us fix the coefficient $h$, and hence the radius, as well as the volume of $\Torus^4$ in terms of the brane charges of the background. We are then left with the 20 expected moduli.

\section{The TrT transformed Green-Schwarz string}
\label{sec:trt-transformed-GS}

In this appendix we write down the expressions for the terms in the gauge-fixed Green-Schwarz string after two TrT transformations with parameters $\varphi$ and $\psi$. In general, the Lagrangian can be written as
\begin{equation}
  \mathcal{L} =-\frac{1}{2} \Bigl(
  \gamma^{\alpha\beta}\partial_\alpha x^i\partial_\beta x^j G_{ij}
  -\epsilon^{\alpha\beta}\partial_\alpha x^i\partial_\beta x^j B_{ij}
  +2\partial_\alpha x^i
  \bigl(\gamma^{\alpha\beta}U_{\beta,i}-\epsilon^{\alpha\beta}V_{\beta,i}\bigr)+\mathcal{L}_{\text{rest}}\Bigr).
\end{equation}
We will work to quadratic order in fermions and decompose the coefficients in the Lagrangian as
\begin{equation}
  G_{ij} = G_{ij}^{(b)} + G_{ij}^{(f)} , \quad
  B_{ij} = B_{ij}^{(b)} + B_{ij}^{(f)} , \quad
  U_{\alpha,i} = U_{\alpha,i}^{(b)} + U_{\alpha,i}^{(f)} , \quad
  V_{\alpha,i} = V_{\alpha,i}^{(b)} + V_{\alpha,i}^{(f)} ,
\end{equation}
where, \eg, $G_{ij}^{(b)}$ is purely bosonic while $G_{ij}^{(f)}$ is quadratic in fermions. For concreteness we will assume that there are four $\grpU(1)$ directions, which are parametrised with $x^6, \dotsc, x^9$. We further set the volume of these four directions to be equal and assume that there is no B field and that the metric is diagonal. The bosonic background fields then take the form
\begin{equation}
  G^{(b)}_{ij} = \delta_{ij} \sqrt{\frac{\mu_1}{\mu_5}} , \qquad
  B_{ij}^{(b)} = 0 , \qquad
  U_{\alpha,i}^{(b)} = 0 , \qquad
  V_{\alpha,i}^{(b)} = 0 .
\end{equation}
Furthermore, we set
\begin{equation}
  G_{ij}^{(f)} = \delta_{ij} g_{f} \sqrt{\frac{\mu_1}{\mu_5}} , \qquad
  B_{89}^{(f)} = B_{67}^{(f)} , \qquad
  B_{79}^{(f)} = -B_{68}^{(f)} , \qquad
  B_{78}^{(f)} = B_{69}^{(f)} .
\end{equation}
Performing the two TrT transformations and keeping only terms that are at most quadratic in fermions, we find that the bosonic part of the resulting background is given by
\begin{equation}
  \begin{aligned}
    \tilde{G}_{66}^{(b)} &= \tilde{G}_{77}^{(b)} = \frac{\sqrt{\mu_1\mu_5}}{\mu_{\varphi}} , \quad &
    \tilde{B}_{67}^{(b)} &= \frac{\mu_1 - \mu_5}{\mu_{\varphi}} \sin\varphi \cos\varphi ,
    \\
    \tilde{G}_{88}^{(b)} &= \tilde{G}_{99}^{(b)} = \frac{\sqrt{\mu_1\mu_5}}{\mu_{\psi}} , &
    \tilde{B}_{89}^{(b)} &= \frac{\mu_1 - \mu_5}{\mu_{\psi}} \sin\psi \cos\psi .
  \end{aligned}
\end{equation}
The terms in the transformed background that are quadratic in fermions are given by
\begin{equation}
  \begin{gathered}
    \tilde{G}_{66}^{(f)} = \tilde{G}_{77}^{(f)} = \frac{\sqrt{\mu_1 \mu_5}}{\mu_{\varphi}^2} \bigl(
    ( \mu_5 \cos^2\varphi - \mu_1 \sin^2\varphi ) g_f - 2 \mu_5 \cos\varphi \sin\varphi \, B_{67}^{(f)} 
    \bigr) ,
    \\
    \tilde{G}_{88}^{(f)} = \tilde{G}_{99}^{(f)} = \frac{\sqrt{\mu_1 \mu_5}}{\mu_{\psi}^2} \bigl(
    ( \mu_5 \cos^2\psi - \mu_1 \sin^2\psi ) g_f - 2 \mu_5 \cos\psi \sin\psi \, B_{67}^{(f)} 
    \bigr) ,
    \\
    \mathllap{-}\tilde{G}_{79}^{(f)} = \tilde{G}_{68}^{(f)} = \frac{\sqrt{\mu_1 \mu_5}}{\mu_{\varphi}\mu_{\psi}} \mu_5 \sin(\varphi-\psi) \, B_{69}^{(f)} ,
    \\
    \tilde{G}_{78}^{(f)} = \tilde{G}_{69}^{(f)} = \frac{\sqrt{\mu_1 \mu_5}}{\mu_{\varphi}\mu_{\psi}} \mu_5 \sin(\varphi-\psi) \, B_{68}^{(f)} ,
    \\
    \tilde{B}_{67}^{(f)} = \frac{\mu_5}{\mu_{\varphi}^2} \bigl(
    ( \mu_5 \cos^2\varphi - \mu_1 \sin^2\varphi ) B_{67}^{(f)} + 2 \mu_1 \cos\varphi \sin\varphi \, g_f
    \bigr)
    \\
    \tilde{B}_{89}^{(f)} = \frac{\mu_5}{\mu_{\psi}^2} \bigl(
    ( \mu_5 \cos^2\psi - \mu_1 \sin^2\psi ) B_{67}^{(f)} + 2 \mu_1 \cos\psi \sin\psi \, g_f
    \bigr) ,
    \\
    \mathllap{-} \tilde{B}_{79}^{(f)} = \tilde{B}_{68}^{(f)} = \frac{\mu_5}{\mu_{\varphi}\mu_{\psi}} ( \mu_5 \cos\varphi \cos\psi + \mu_1 \sin\varphi \sin\psi ) B_{68}^{(f)} ,
    \\
    \tilde{B}_{78}^{(f)} = \tilde{B}_{69}^{(f)} = \frac{\mu_5}{\mu_{\varphi}\mu_{\psi}} ( \mu_5 \cos\varphi \cos\psi + \mu_1 \sin\varphi \sin\psi ) B_{69}^{(f)} ,
    \\
    \tilde{U}_{\alpha,6}^{(f)} = \frac{\mu_5}{\mu_\varphi} \cos\varphi \, U_{\alpha,6}^{(f)} + \frac{\sqrt{\mu_1\mu_5}}{\mu_\varphi} \sin\varphi \, V_{\alpha,7}^{(f)} ,
    \\
    \tilde{U}_{\alpha,7}^{(f)} = \frac{\mu_5}{\mu_\varphi} \cos\varphi \, U_{\alpha,7}^{(f)} - \frac{\sqrt{\mu_1\mu_5}}{\mu_\varphi} \sin\varphi \, V_{\alpha,6}^{(f)} ,
    \\
    \tilde{U}_{\alpha,8}^{(f)} = \frac{\mu_5}{\mu_\psi} \cos\psi \, U_{\alpha,8}^{(f)} + \frac{\sqrt{\mu_1\mu_5}}{\mu_\psi} \sin\psi \, V_{\alpha,8}^{(f)} ,
    \\
    \tilde{U}_{\alpha,9}^{(f)} = \frac{\mu_5}{\mu_\psi} \cos\psi \, U_{\alpha,9}^{(f)} - \frac{\sqrt{\mu_1\mu_5}}{\mu_\psi} \sin\psi \, V_{\alpha,9}^{(f)} ,
    \\
    \tilde{V}_{\alpha,6}^{(f)} = \frac{\mu_5}{\mu_\varphi} \cos\varphi \, V_{\alpha,6}^{(f)} + \frac{\sqrt{\mu_1\mu_5}}{\mu_\varphi} \sin\varphi \, U_{\alpha,7}^{(f)} ,
    \\
    \tilde{V}_{\alpha,7}^{(f)} = \frac{\mu_5}{\mu_\varphi} \cos\varphi \, V_{\alpha,7}^{(f)} - \frac{\sqrt{\mu_1\mu_5}}{\mu_\varphi} \sin\varphi \, U_{\alpha,6}^{(f)} ,
    \\
    \tilde{V}_{\alpha,8}^{(f)} = \frac{\mu_5}{\mu_\psi} \cos\psi \, V_{\alpha,8}^{(f)} + \frac{\sqrt{\mu_1\mu_5}}{\mu_\psi} \sin\psi \, U_{\alpha,8}^{(f)} ,
    \\
    \tilde{V}_{\alpha,9}^{(f)} = \frac{\mu_5}{\mu_\psi} \cos\psi \, V_{\alpha,9}^{(f)} - \frac{\sqrt{\mu_1\mu_5}}{\mu_\psi} \sin\psi \, U_{\alpha,9}^{(f)} .
  \end{gathered}
\end{equation}

\bibliographystyle{nb}
\bibliography{refs,FI-paper}

\end{document}